\documentclass[%
 twocolumn,
 longbibliography,
superscriptaddress,
 amsmath,amssymb,
 aps,
pra,
]{revtex4-2}

\usepackage{hyperref}

\usepackage{graphicx}% Include figure files
\usepackage{dcolumn}% Align table columns on decimal point
\usepackage{bm}% bold math

\usepackage{ulem}
\usepackage{color}
\newcommand{\wzy}[1]{\textcolor{black}{#1}}

\begin{document}

\title{Wide-band Unambiguous Quantum Sensing via Geodesic Evolution}

\author{Ke Zeng}
\affiliation{Key Laboratory of Atomic and Subatomic Structure and Quantum Control
(Ministry of Education), and School of Physics, South China Normal
University, Guangzhou 510006, China}

\author{Xiaohui Yu}
\affiliation{Key Laboratory of Atomic and Subatomic Structure and Quantum Control
(Ministry of Education), and School of Physics, South China Normal
University, Guangzhou 510006, China}

\author{Martin B. Plenio}
\affiliation{Institut für Theoretische Physik und IQST, Universität Ulm, 
Albert-Einstein-Allee 11, 89081 Ulm, Germany}

\author{Zhen-Yu Wang}
\email{zhenyu.wang@m.scnu.edu.cn}
\affiliation{Key Laboratory of Atomic and Subatomic Structure and Quantum Control
(Ministry of Education), and School of Physics, South China Normal
University, Guangzhou 510006, China}
\affiliation{Guangdong Provincial Key Laboratory of Quantum Engineering and
Quantum Materials, and Guangdong-Hong Kong Joint Laboratory of Quantum
Matter, South China Normal University, Guangzhou 510006, China}

\begin{abstract}
We present a quantum sensing technique that utilizes a sequence of $\pi$ pulses to cyclically drive the qubit dynamics along a geodesic path of adiabatic evolution. This approach effectively suppresses the effects of both decoherence noise and control errors while simultaneously removing unwanted resonance terms, such as higher harmonics and spurious responses commonly encountered in dynamical decoupling control. As a result, our technique offers robust, wide-band, unambiguous, and high-resolution quantum sensing capabilities for signal detection and individual addressing of quantum systems, including spins. To demonstrate its versatility, we showcase successful applications of our method in both low-frequency and high-frequency sensing scenarios. The significance of this quantum sensing technique extends to the detection of complex signals and the control of intricate quantum environments. By enhancing detection accuracy and enabling precise manipulation of quantum systems, our method holds considerable promise for a variety of practical applications.
\end{abstract}
\maketitle

\paragraph*{Introduction. ---}

Accurate characterization of the qubit environment holds significant importance across a range of 
applications, spanning from quantum information processing to quantum sensing 
\citep{preskill2018quantum,paladino2014,degen2017quantum,barry2020sensitivity}. A widely utilized 
technique for achieving this is the implementation of dynamical decoupling (DD) pulse sequences
\citep{viola1999dynamical,yang2011preserving}. These sequences serve to filter out environmental 
noise, thereby extending the quantum coherence time, as well as to extract and amplify signals of 
specific frequencies \citep{degen2017quantum,barry2020sensitivity,javaloy2021dynamical}. 
Consequently, qubits under such sequences become highly sensitive quantum sensors, presenting diverse applications.
For instance, nitrogen-vacancy (NV) centers \citep{doherty2013the,WuJPW16,weber2010quantum} subjected
to DD pulse sequences enable nanoscale nuclear magnetic resonance (NMR) \citep{kolkowitz2012sensing,zhao2012sensing,staudacher2013nuclear,muller2014nuclear,rugar2015,deVience2015nanoscale,
glenn2018high,lang2017enhanced,pfender2019high,bucher2019quantum,casanova2019modulated,aharon2019quantum,
javaloy2020robust,meinel2022quantum,wang2023using,javaloy2023high,javaloy2021dynamical},
spin label detection \citep{shi2015single,lovchinsky2016nuclear,javaloy2022detection},
spin cluster imaging \citep{zhao2011atomic,shi2014sensing,lang2015dynamical,wang2016positioning,sasaki2018determination,zopes2018three,zopes2018threePRL,abobeih2019atomic,cujia2022parallel},
and AC field sensing \citep{schmitt2017submillihertz,boss2017quantum,joas2017quantum,stark2017narrow,chu2021precise,meinel2021heterodyne,wang2022sensing,jiang2023quantum}.
Furthermore, they can be employewd for controlling nearby single nuclear spins \citep{taminiau2012detection,wang2017delayed,haase2018soft,perlin2019noise,hegde2020efficient,bartling2022entanglement}
in the context of quantum information processing \citep{casanova2016noise,pezzagna2021quantum},
quantum simulations \citep{cai2013a}, and quantum networks \citep{childress2013diamond,kalb2017entanglement,humphreys2018deterministic,pompili2021realization}.

\begin{figure}
\includegraphics[width=1\columnwidth]{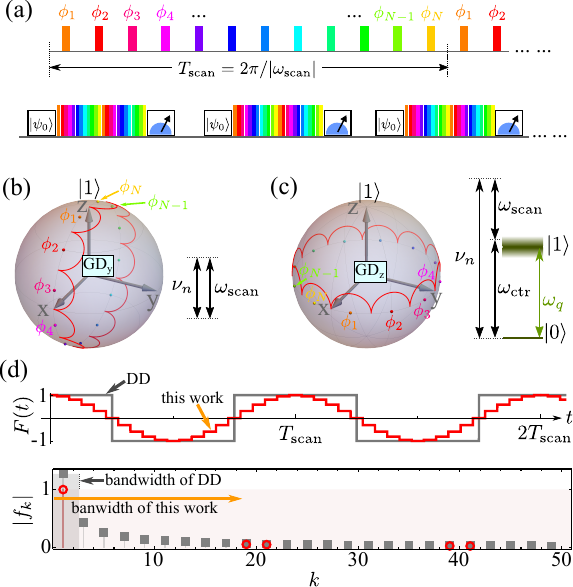}\caption{\label{fig:Scheme} Quantum sensing via geodesic 
jumping. (a) Upper
panel: Repeated application of a sequence of $N$ $\pi$ pulses realizes
cyclic quantum adiabatic evolution along the geodesic in (b) or (c).
Lower panel: Combined with synchronized readout techniques for arbitrary
frequency resolution. (b) \wzy{$\text{GD}_{y}$}, where the closed geodesic is
sampled by $N$ $\pi$ pulses. The red solid line illustrates the trajectory
of the state evolution starting at $|1\rangle$ for $N=12$. When $\omega_{{\rm scan}}$ matches the 
frequency $\nu_{n}$ of the target, a resoance occurs. (c) As
(b) but for \wzy{$\text{GD}_{z}$} which uses a horizontal geodesic, e.g., for
robust heterodyne sensing of high-frequency signal. The resonant condition
is accurately tuned by $T_{{\rm scan}}$ and the frequency $\omega_{{\rm ctr}}$
of control field. (d) The resulting modulation function
$F(t)$ (red line for $N=20$) and its Fourier components (red cycles)
where $f_{k}=0$ for all $1<k<N-1$. The gray line and squares are
the corresponding ones for DD pulse sequences. }
\end{figure}

However, DD pulse sequences used for quantum sensing, such as the commonly employed 
Carr-Purcell-Meiboom-Gill (CPMG) \citep{carr1954effects,meiboom1958modified} and XY8 
\citep{gullion1990new} sequences encounter an issue known as spectral leakage and spurious 
resonance. These complications make the interpretation of the sensor's recorded signal challenging 
and can result in ambiguous signal identification \citep{loretz2015spurious,frey2017application}.
DD pulses introduce abrupt temporal state flipping of the qubit sensor, causing a pronounced 
frequency modulation at a specific frequency [indicated by the gray line in Fig.~\ref{fig:Scheme}(d){]}.
This modulation leads to a strong resonance at the flipping frequency, enabling frequency-selective 
sensing. However, it also generates resonances at other frequencies, as evident in its Fourier 
transform [represented by the gray squares in Fig.~\ref{fig:Scheme}(d){]}.
Consequently, signals outside the target sensing frequency band can contribute to the sensor's 
response and each individual frequency signal produces multiple signal peaks in the 
measured spectrum. Complex signals with a wide frequency range often exhibit significant signal 
overlap, posing challenges for analysis. Additionally, quantum heterodyne methods employed to 
down-convert high-frequency signals for sensing can exacerbate signal overlap 
\citep{wang2022sensing,meinel2021heterodyne,chu2021precise}. Furthermore, the limited power of 
control pulses introduces spurious signals at unexpected frequencies \citep{loretz2015spurious}.
These factors collectively present obstacles to reliably measuring environmental signals, particularly 
when various background noise sources are not fully characterized. With this goal in mind sequence 
timing has been optimised ~\citep{zhao2014dynamical,casanova2015robust} to eliminate 
higher-harmonics and sequence randomization has been explored to mitigates spurious signals
\citep{wang2019randomization,wang2020enhancing}. However, both approaches can address only specific 
their aspects and can solve these only partially.

In this Letter, we present a novel approach called cyclic geodesic driving, which enables 
unambiguous sensing of signals across a broad frequency range on qubit sensors. Our method 
involves the application of a sequence of $\pi$-pulses to implement accelerated high-fidelity quantum 
adiabatic driving, effectively inducing periodic evolution of the quantum sensor along the geodesic 
path on the Bloch sphere. As a result, the resonance frequency of the quantum sensor aligns with the 
frequency of the periodic evolution, while simultaneously mitigating the impact of environmental 
background noise.
By employing this technique, each individual frequency signal generates a single, distinct signal 
response within the wide frequency band. This eliminates undesirable signal overlap, enabling 
precise characterization of complex external environments and signals. For achieving arbitrary 
high frequency resolution, our method can be combined with synchronized readout techniques
\citep{schmitt2017submillihertz,boss2017quantum}. Furthermore, our approach exhibits remarkable 
resilience against control errors, as it is the counterpart of quantum adiabatic control.
Overall, our proposed method of cyclic geodesic driving offers a robust and effective solution 
for wide-band, unambiguous signal sensing, enabling accurate analysis of intricate quantum systems 
and external environments.

\paragraph*{Adiabatic shortcut by jumping. ---}

Our sensing scheme utilizes a sequence of periodic $\pi$ pulses {[}Fig.
1 (a){]} which achieves cyclic quantum adiabatic evolution along a geodesic 
\citep{wang2016necessary,xu2019breaking,liu2022shortcuts,gong2023accelerated} defined by the 
control Hamiltonian for the qubit sensor 
\begin{equation}
H_{c}(t)=\frac{E(t)}{2}(|+_{\phi}\rangle\langle+_{\phi}|-|-_{\phi}\rangle\langle-_{\phi}|),\label{eq:Hc}
\end{equation}
where the instantaneous eigenstates $|\pm_{\phi}\rangle$ are varied by a %monotonic 
parameter $\phi=\phi(t)$ starting with $\phi(0)=0$ at the initial time $t=0$. 
For adiabatic evolution of the parameter $\phi(t)$, the evolution operator 
for the quantum state driven by $H_{c}(t)$ reads \citep{berry2009transitionless,wang2016necessary,odelin2019shortcuts}
\begin{equation}
	U_{c}(\phi)=e^{-i\varphi_{+}(t)}|+_{\phi}\rangle\langle+_{0}|+e^{-i\varphi_{-}(t)}|-_{\phi}\rangle\langle-_{0}|,\label{eq:Uc}
\end{equation}
which transfers the initial state $|\pm_{0}\rangle$ to the instantaneous eigenstate $|\pm_{\phi}\rangle$ 
at a later time. We use the Born-Fock gauge $\langle\pm_{\phi}|\frac{d}{d\phi}|\pm_{\phi}\rangle=0$
such that $\varphi_{\pm}(t)=\pm\frac{1}{2}\int_{0}^{t}E(t^{\prime})dt^{\prime}$
are the dynamic phases. To realize the adiabatic evolution $U_{c}(\phi)$
by $H_{c}(t)$ in finite times, we choose
\begin{equation}
\begin{aligned}
|+_{\phi}\rangle= & \cos(\frac{\phi}{2})|+_{0}\rangle+\sin(\frac{\phi}{2})|-_{0}\rangle,\\
|-_{\phi}\rangle= & -\sin(\frac{\phi}{2})|+_{0}\rangle+\cos(\frac{\phi}{2})|-_{0}\rangle,
\end{aligned}\label{eq:geodesics}
\end{equation}
which connect two orthonormal states $|\pm_{0}\rangle$ via a geodesic
curve \citep{chruscinski2004geometric}. In our scheme we apply a
sequence of control $\pi$ pulses
via Eq.~(\ref{eq:Hc}) with $\phi=\phi_{j}$ at the moments $T_{j}=T_{{\rm scan}}\frac{2j-1}{2N}$
($j=1,2,\ldots$), where $N$ is the pulse number in one periodic
of evolution, see Fig.~\ref{fig:Scheme}(a). We use the linear relation $\phi_{j}=\omega_{{\rm scan}}T_{j}$,
where frequency $\omega_{{\rm scan}}$
can be negative or positive depending on the change of $\phi_{j}$
in time. Each control $\pi$ pulse has a time duration $t_{j}$ such
that the pulse area $\int_{T_{j}-t_{j}/2}^{T_{j}+t_{j}/2}E(t)dt=\pi$.
Between the $\pi$ pulses, there is no control, i.e., $E(t)=0$ if
$\phi\notin\{\phi_{j}\}$. We remove the dynamic phases at the final
time of the evolution, we introduce a $\pi$ phase shift to the pulses
in the second-half of the sequence. According to Refs. \citep{wang2016necessary,xu2019breaking,liu2022shortcuts},
the sequence realizes $U_{c}(\phi)$ with unit-fidelity at the middle
of any successive path points $\phi=\bar{\phi}_{j}\equiv(\phi_{j}+\phi_{j-1})/2$. 
For other values of $\phi\notin\{\bar{\phi}_{j}\}$
the difference between $U_{c}(\phi)$ and the actual evolution implemented
by $H_{c}(t)$ is negligible when $N$ is sufficiently large. In contrast
to conventional shortcuts to adiabaticity~\citep{odelin2019shortcuts} to accerate adiabatic process,
in our method the instantaneous eigenstates of the control Hamiltonian
$H_{c}(t)$ are the same as the evolution path $|\pm_{\phi}\rangle$ in Eq.~(\ref{eq:Uc}).
This avoids the use of counter-diabatic fields and retains the intrinsic
robustness of traditional adiabatic process \citep{liu2022shortcuts}.

\begin{figure}
\includegraphics[width=1\columnwidth]{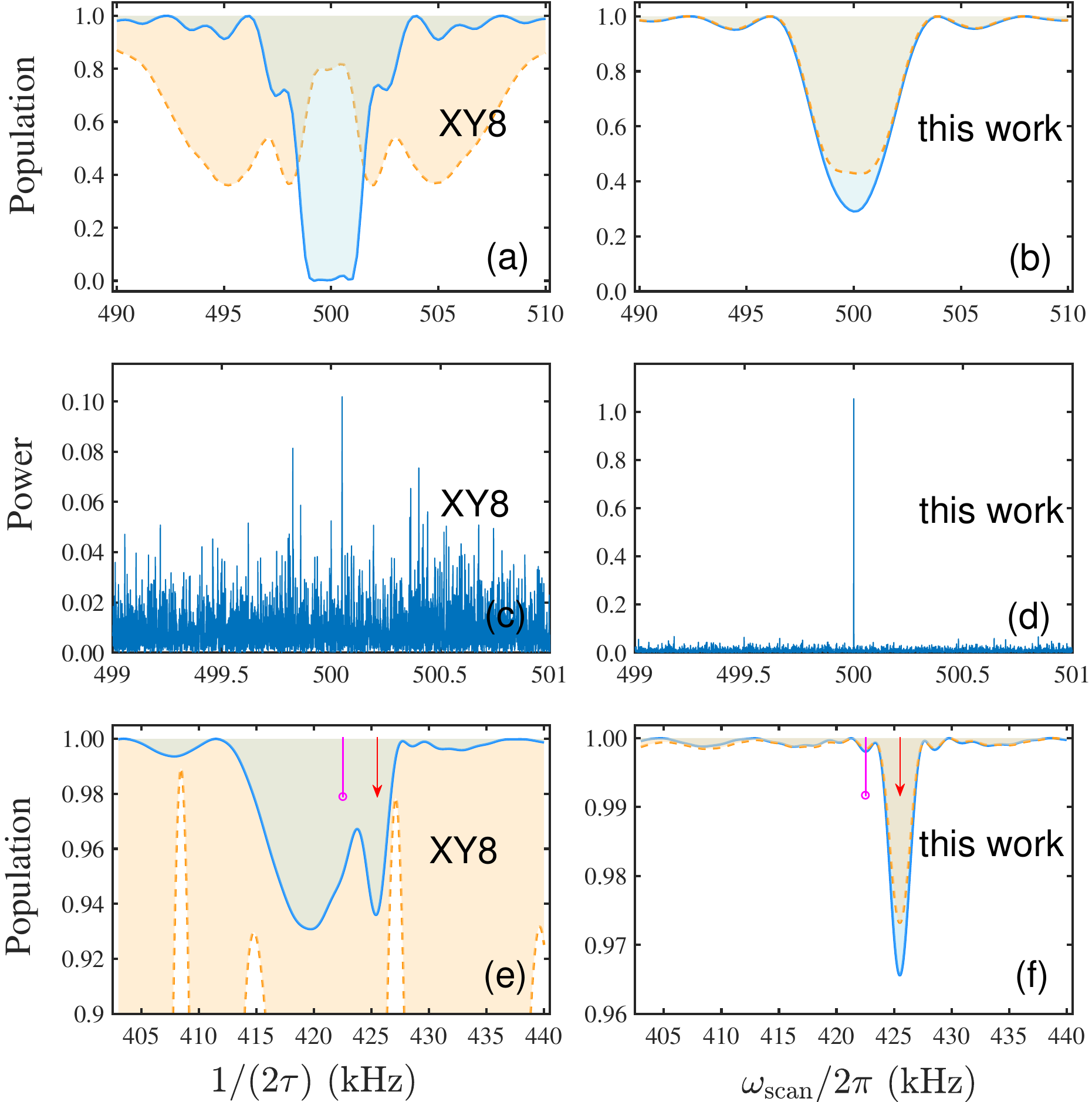}\caption{\label{fig:ZX} Quantum spectroscopy. (a) Population signal of XY8
sequence (blue solid line) for spectroscopy of AC fields with the
frequencies $\{\nu_{j}/2\pi\}=\{500,1500.05,2499.88\}$ kHz by varying
the pulse interval $\tau$. The 1500.05 kHz and 2499.88 kHz AC fields
distort the signal centered at 500 kHz via the 3rd and 5th harmonics,
respectively. Yellow dashed line is the result when there is dephasing
noise (with a control-free decoherence time $T_{2}^{*}\approx2$ $\mu$s)
and control errors (with about $20$\% drift on the amplitude of control
field). (b) As (a) but by using \wzy{$\text{GD}_{y}$}. The 500 kHz signal dip
is not distorted by the 1500.05 kHz and 2499.88 kHz AC fields. (c)
Power spectrum for the AC signal fields in (c) by using the synchronized
readout technique in Ref. \citep{boss2017quantum,schmitt2017submillihertz}.
The resonances due to higher harmonics make the signal unidentifiable
even through the expected spectral resolution is about 1 Hz. (d) As
(c) but by using \wzy{$\text{GD}_{y}$}. (e) Signal of XY8 sequence for the detection
of a $^{1}\text{H}$ spin with its frequency indicated by a red arrow.
The spurious resonance (centered around the pink vertical line) produced
by a $^{13}\text{C}$ spin in diamond distorts the $^{1}\text{H}$
spin signal. Yellow dashed line is the result when there are $2\pi\times2$
MHz detuning error and 30\% of amplitude drift in the control field.
(f) As (e) but by using \wzy{$\text{GD}_{y}$}, where the spurious signals disappear. See \citep{see_SM}for details
of simulation. }
\end{figure}

\paragraph*{
\begin{figure}
\protect\includegraphics[width=1\columnwidth]{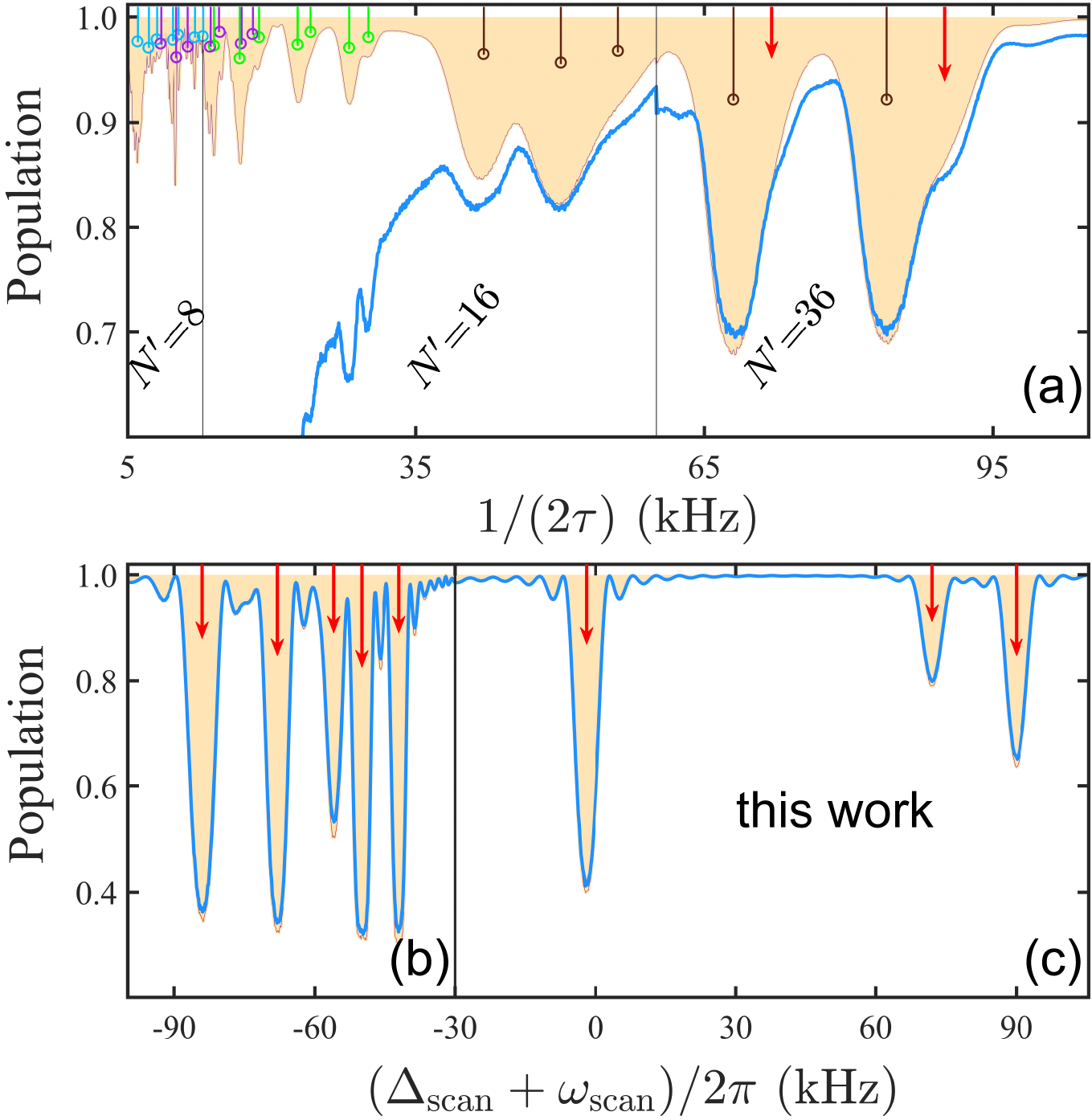}\protect\caption{\label{fig:XY} (a) Quantum heterodyne spectroscopy of a signal field
with frequencies $\{\nu_{j}\}=\omega_{q}+2\pi\times\{-84,-68,-56,-50,-42,-2,72,90\}$
kHz by using the protocol in Refs. \citep{chu2021precise,meinel2021heterodyne}
with the CPMG sequences. $\omega_{q}$ could be at the GHz range.
The blue solid line (the line with yellow filling) is the simulation
with (without) dephasing noise that induces a control-free decoherence
time $T_{2}^{*}\approx2$ $\mu$s. The pulse interval $\tau$ is varied
to measure the spectrum. Because for each AC signal frequency
$\nu_{j}$ resonance dips occur whenever $1/(2\tau)=(\nu_{j}-\omega_{{\rm ctr}})/(2\pi k)$,
($k=\pm 1,\pm 3, \pm 5,\ldots$), the dips at $1/(2\tau)=(\nu_{j}-\omega_{q})/(2\pi)$
indicated by red vertical arrows are obscured by other resonance dips
(vertical lines). Different number $N^{\prime}$ of $\pi$
pulses are used for different range of $\tau$ to insure that the
sequence times are smaller than 1 ms. (b) {[}(c){]}: As (a) but by using Protocol with
fixed $\Delta_{{\rm scan}}=0$ ($\omega_{{\rm scan}}=2\pi\times80$
kHz{]}). All the dips only appear at the right frequencies $\nu_j$. See \citep{see_SM} for details of simulation. }
\end{figure}
}

\paragraph*{Unambiguous wide-bandwidth robust sensing. ---}

To demonstrate the concept of quantum sensing through the aforementioned
geometric control, we examine a qubit coupled to its environment
via the interaction Hamiltonian 
\begin{equation}
H_{{\rm int}}(t)=\frac{1}{2}\sigma_{z}B(t),
\end{equation}
where the Pauli operator $\sigma_{z}=|1\rangle\langle1|-|0\rangle\langle0|$
and $B(t)$ could be a classical field or a quantum operator in a
rotating frame which includes possible dephasing noise.  We use
the control Hamiltonian $H_{c}(t)=\Delta(t)\frac{\sigma_{z}}{2}+\Omega(t)\frac{\sigma_{x}}{2}$
and the states $|+_{0}\rangle=|1\rangle$ and $|-_{0}\rangle=|0\rangle$
for the geodesic in Eq.~(\ref{eq:geodesics}) for sensing. 
\wzy{This geodesic driving around the $y$-axis ($\text{GD}_{y}$)} is sketched in Fig.~\ref{fig:Scheme}(b).
In the rotating frame of $H_{c}(t)$, $H_{{\rm int}}(t)$ becomes
$\tilde{H}_{{\rm int}}(t)\approx\frac{1}{2}U_{c}^{\dagger}\sigma_{z}U_{c}B(t)$.
For the evolution Eq.~(\ref{eq:Uc}), we find the approximated transformation
\citep{see_SM} 
%\begin{align}
%U_{c}^{\dagger}\sigma_{z}U_{c} & \approx\cos\phi\sigma_{z},\nonumber \\
%U_{c}^{\dagger}\sigma_{x}U_{c} & \approx\sin\phi\sigma_{z},\label{eq:PauliTranformedApprox}\\
%U_{c}^{\dagger}\sigma_{y}U_{c} & \approx0,\nonumber 
%\end{align}
\begin{equation}
U_{c}^{\dagger}\sigma_{z}U_{c} \approx\cos\phi\sigma_{z};~U_{c}^{\dagger}\sigma_{x}U_{c} \approx\sin\phi\sigma_{z};~
U_{c}^{\dagger}\sigma_{y}U_{c} \approx0,\label{eq:PauliTranformedApprox}
\end{equation}
which have the nice property that they only depend on the geometric paramteter $\phi(t)$. 
We obtain \citep{see_SM}
\begin{equation}
\tilde{H}_{{\rm int}}(t)\approx F(t)\frac{\sigma_{z}}{2}B(t),\label{eq:HintV}
\end{equation}
where when $N$ is sufficiently large the modulation function $F(t)\approx\cos(\omega_{{\rm scan}}t)$
has only one Fourier component over a large frequency band, see Fig.~\ref{fig:Scheme}(d). 
Conventional DD sequences \citep{degen2017quantum,barry2020sensitivity}
also induce modulation factors $F(t)$ to the $\sigma_{z}$ operator
with $F(t)\in\{\pm1\}$ for ideal sequences \citep{cywinski2008how,degen2017quantum,barry2020sensitivity}.
However, those modulation factors have multiple Fourier components
that complicate the interpretation of the sensor's signal \citep{zhao2014dynamical,loretz2015spurious,frey2017application},
see Fig.~\ref{fig:Scheme}(d) for equally spaced DD sequences \citep{degen2017quantum,barry2020sensitivity,carr1954effects,meiboom1958modified,gullion1990new,alvarez2008measuring,genov2017arbitrarily,ryan2010robust}. 

\begin{figure}
\includegraphics[width=1\columnwidth]{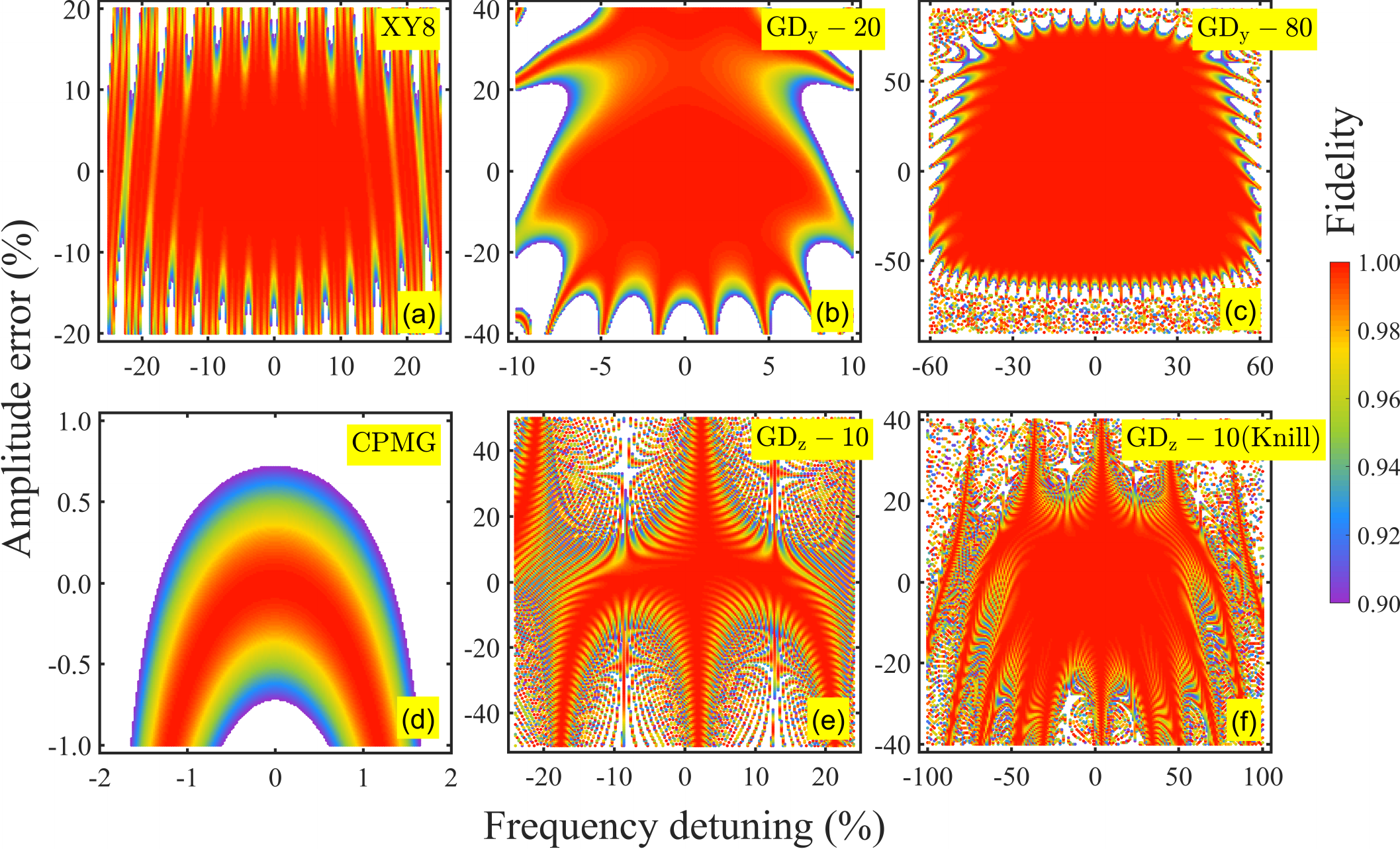}\caption{\label{fig:Robust} Control fidelity with respect to amplitude and
detuning errors for (a) XY8 sequence, (b) \wzy{$\text{GD}_{y}$} with $N=20$,
(c)\wzy{$\text{GD}_{y}$} with $N=80$,  (d) CPMG
sequence with $40$ pulses for heterodyne sensing, (e) \wzy{$\text{GD}_{z}$}
with $N=10$, (f) \wzy{$\text{GD}_{z}$} with $N=10$ but each pulse is replaced
by a Knill pulse. All the protocols have the same sequence time length,
and the ideal Rabi frequency of the control is $2\pi\times50$ MHz.}
\end{figure}

In Figs. \ref{fig:ZX}(a)-(d) we simulate the measured spectrum of
a classical AC field with $B(t)=\sum_{j=1}^{3}b_{j}\cos(\nu_{j}t+\theta_{j})$,
where $\nu_{j}$ are the frequencies of different components.  For
the result of Fig.~\ref{fig:ZX}(a) obtained by the widely-used robust
XY8 sequence with an interpulse duration $\tau$ \citep{gullion1990new},
all the frequencies ($\{\nu_{j}/2\pi\}=\{500,1500.05,2499.88\}$ kHz)
cause transitions of the sensor states at $1/(2\tau)\approx500$ kHz
via the 1st, 3rd, and the 5th harmonics. This problem of ambiguous
spectral overlap is not solved even when we improve the frequency
resolution sufficiently high via the synchronized readout technique
(Qdyne) \citep{schmitt2017submillihertz,boss2017quantum}, see Fig.~\ref{fig:ZX}(c), because all the frequency components contribute
to the readout signal in Qdyne. However, using \wzy{$\text{GD}_{y}$}, only the
frequency $\nu_{1}/2\pi=500$ kHz contributes to the dip at the resonance
$\omega_{{\rm scan}}=\nu_{1}$, because we have the effective Hamiltonian
$\tilde{H}_{{\rm int}}\approx\frac{1}{4}b_{1}\cos\theta_{1}\sigma_{z}$
from Eq.~(\ref{eq:HintV}) after rotating wave approximation
\citep{see_SM}. The phase dependence on the effective Hamiltonian
(and hence the signal) allows for arbitrary frequency resolution with 
synchronized readout, see Figs. \ref{fig:ZX}(d) and \ref{fig:Scheme}(a). 
The results also show that our protocol is more resistant to control 
errors and dephasing noise. The already strong robustness of 
\wzy{$\text{GD}_{y}$} is enhanced further with increasing number $N$ of
pulses, see Fig.~\ref{fig:Robust}.

It is interesting that \wzy{cyclic geodesic driving} also fully
solve the problems of spurious response due to finite-width pulses
\citep{loretz2015spurious}, as detailed in \citep{see_SM}. In Fig.~\ref{fig:ZX}(e) and (f), we simulate the quantum sensing of a single
proton spin ($^{1}{\rm H}$) by an NV center. A $^{13}{\rm C}$ spin
in diamond is also coupled to the NV center as a noise source.
For this case $B(t)$ is a quantum operator \citep{see_SM,wang2017delayed,haase2018soft}.  
For the result Fig.~\ref{fig:ZX}(e) obtained
by XY8 sequences, the spurious response from the $^{13}{\rm C}$ spin
disturb the target proton spin signal and lead to misidentification
of $^{13}{\rm C}$ nuclei for proton. On the contrary, \wzy{$\text{GD}_{y}$}
provides a clean signal dip in the spectrum {[}Fig.~\ref{fig:ZX}(f){]}, 
because when $\omega_{{\rm scan}}$ matches the frequency
$\nu_{1}$ of the target $^{1}{\rm H}$, $\tilde{H}_{{\rm int}}\approx\frac{1}{2}a_{1}^{x}\sigma_{z}I_{1}^{x}$ (where $I_{1,z}$ is the $^{1}{\rm H}$ spin operator and $a_{1}^{x}$ is the hyperfine strength  \citep{see_SM}) 
and the effect of the $^{13}{\rm C}$ spin is removed. 

\paragraph*{Unambiguous heterodyne sensing. ---}

The idea can be generalized to other settings. Consider the sensing
of a multi-frequency signal field $\vec{B}(t)=(B_{x},B_{y},B_{z})$
with frequencies $\nu_{j}$ much larger than the Rabi frequency of
the control field. We use \wzy{$\text{GD}_{z}$} {[}see Fig.~\ref{fig:Scheme}(c){]}
 with $|\pm_{0}\rangle=(|0\rangle\pm|1\rangle)/\sqrt{2}$ for
the geodesic in Eq.~(\ref{eq:geodesics}) to sense the traverse part
$B_{\perp}(t)\equiv B_{x}+iB_{y}=\sum_{j}b_{j}e^{i\alpha_{j}}\cos(\nu_{j}t+\theta_{j})$.
 The relevant Hamiltonian reads
\[
H=\frac{1}{2}(\omega_{q}+\delta_{t})\sigma_{z}+(\frac{1}{2}B_{\perp}(t)\sigma_{+}+{\rm h.c.})+H_{{\rm ctr}},
\]
where $\omega_{q}$ is the frequency of the qubit and $\delta_{t}$
is an unknown fluctuation. For NV qubits, $\omega_{q}$ (e.g.,
$\omega_{q}=D+\gamma_{e}B_{z}$ with $D=2\pi\times2.87$ GHz and $\gamma_{e}\approx2\pi\times2.8$
MHz/G) can be adjusted by changing the magnetic field $B_{z}$. The
control Hamiltonian $H_{{\rm ctr}}=\Omega(t)\cos(\omega_{{\rm ctr}}t+\phi)\sigma_{x}$
has a controllable detuning $\Delta_{{\rm scan}}=\omega_{{\rm ctr}}-\omega_{q}$.
In the rotating frame with respect to $\frac{1}{2}(\omega_{q}+\delta_{t})\sigma_{z}+H_{{\rm ctr}}$, we obtain the effective
interaction \citep{see_SM} 
\begin{equation}
\tilde{H}_{{\rm int}}(t)\approx\frac{1}{4}\sum_{j}b_{j}e^{i(\alpha_{j}-\theta_{j})+i(\omega_{{\rm ctr}}+\omega_{{\rm scan}}-\nu_{j})t}\sigma_{+}+{\rm h.c.}.
\end{equation}
Under the resonance condition for a signal frequency $\omega_{{\rm ctr}}+\omega_{{\rm scan}}=\nu_{n}$
and when $b_{j}\ll|\nu_{n}-\nu_{j}|$ for $j\neq n$, $\tilde{H}_{{\rm int}}(t)\approx\frac{1}{4}b_{n}e^{i(\alpha_{n}-\theta_{n})}\sigma_{+}+{\rm h.c.}$
picks up the signal only at the frequency $\nu_{n}$. 

As exemplified in Fig.~\ref{fig:XY}(a), the heterodyne sensing using
the CPMG sequences produce multiple dips at $1/(2\tau)=\pm(\nu_{j}-\omega_{q})/k$
$(k=1,3,5,\ldots)$, which implies ambiguity sensor responses especially
when $\{\nu_{j}\}$ happen to be close to the qubit frequency $\omega_{q}$.
In contrast, our geodesic driving gives clear signal dips for accurate
determination of all the signal frequencies, see Fig.~\ref{fig:XY}(b). Our method is also more resilient to dephasing noise (Fig.~\ref{fig:XY}) and is more robust against control errors (Fig.~\ref{fig:Robust}). The robustness can be further enhanced by combining composite pulse techniques
{[}see Fig.~\ref{fig:Robust} (e) for the result where each pulse
is replaced by a Knill pulse \citep{tycko1984fixed,ryan2010robust,casanova2015robust}{]}. 

\paragraph*{Conclusion. ---}
We propose a quantum sensing scheme based on a quantum adiabatic shortcut along 
a geodesic path. This scheme offers the capability to resolve complex broadband signals 
and addresses the issues of spectral overlap and spurious signals that arise in existing 
DD-based quantum sensing methods. Notably, our approach allows for arbitrary frequency 
resolution through the utilization of synchronized readout techniques. Moreover, it 
exhibits robustness against control errors and effectively suppresses unwanted decoherence 
noise. The versatility of our method extends beyond signal detection; it can also be employed 
for the detection and control of various quantum objects, including single nuclear spins, 
spin clusters, and mechanical oscillators. Furthermore, our scheme holds promise for 
applications aimed at mitigating crosstalk in qubit arrays. In summary, our proposed quantum 
sensing scheme based on a quantum adiabatic shortcut along a geodesic path provides a universal 
solution for accurate signal detection, offering improved performance over existing methods. 
Its potential applications encompass a wide range of quantum systems and address key challenges 
in the field.

\begin{acknowledgments}
This work was supported by National Natural Science Foundation of
China (Grant No. 12074131), the Natural Science Foundation of Guangdong 
Province (Grant No. 2021A1515012030), the ERC Synergy grant HyperQ 
(grant no. 856432) and the BMBF Zukunftscluster QSense: Quantensenoren 
f{\"u}r die biomedizinische Diagnostik (QMED) (grant no 03ZU1110FF).
\end{acknowledgments}

\pagebreak{}\clearpage{}\widetext

\begin{center}
\textbf{\large{{{Supplemental Material:\\ Wide-band Unambiguous Quantum Sensing via Geodesic
Evolution}}}}{\large{{ }}} 
\par\end{center}

%%%%%%%%%% Merge with supplemental materials %%%%%%%%%%
%%%%%%%%%% Prefix a "S" to all equations, figures, tables and reset the counter %%%%%%%%%%
\setcounter{equation}{0} \setcounter{figure}{0} \setcounter{table}{0}
\setcounter{page}{1} \makeatletter \global\long\def\theequation{S\arabic{equation}}
 \global\long\def\thefigure{S\arabic{figure}}
 \global\long\def\bibnumfmt#1{[S#1]}
 \global\long\def\citenumfont#1{S#1}
% \tableofcontents{}

\section{Spectral responses }

\subsection{Higher harmonic and spurious responses in DD sequences.}

\begin{figure}
\includegraphics[width=1\columnwidth]{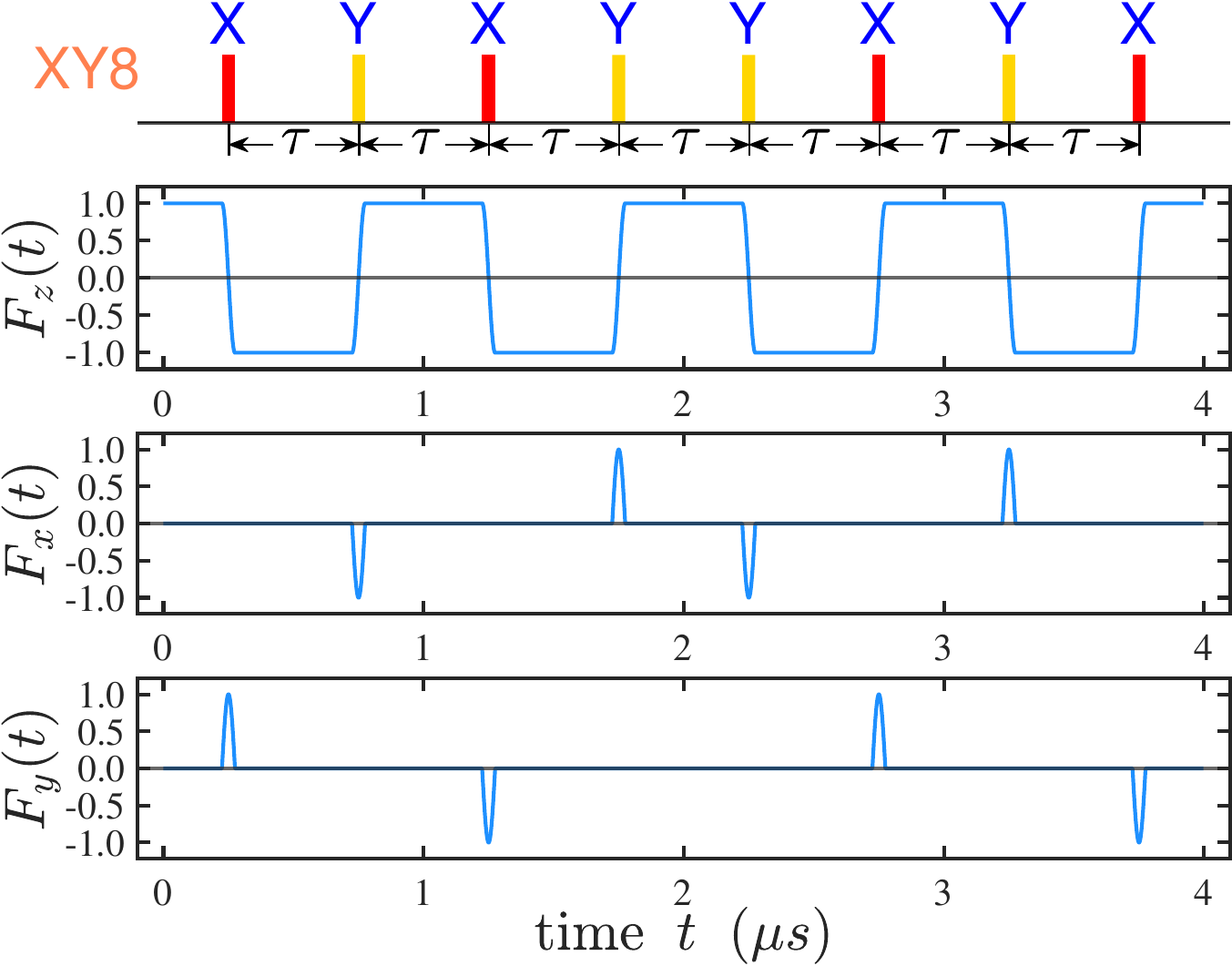}\caption{\label{fig:smXY8} XY8 sequence and its induced modulation functions
$F_{\alpha}(t)$. The $\pi$ pulses of the XY8 sequence have a constant
high $\Omega=2\pi\times10$ MHz and a time interval, $\tau$, between
pulses. The value of $\tau$ determines the frequency $1/(2\tau)=1$
MHz for sensing.}
\end{figure}

\begin{figure}
\includegraphics[width=1\columnwidth]{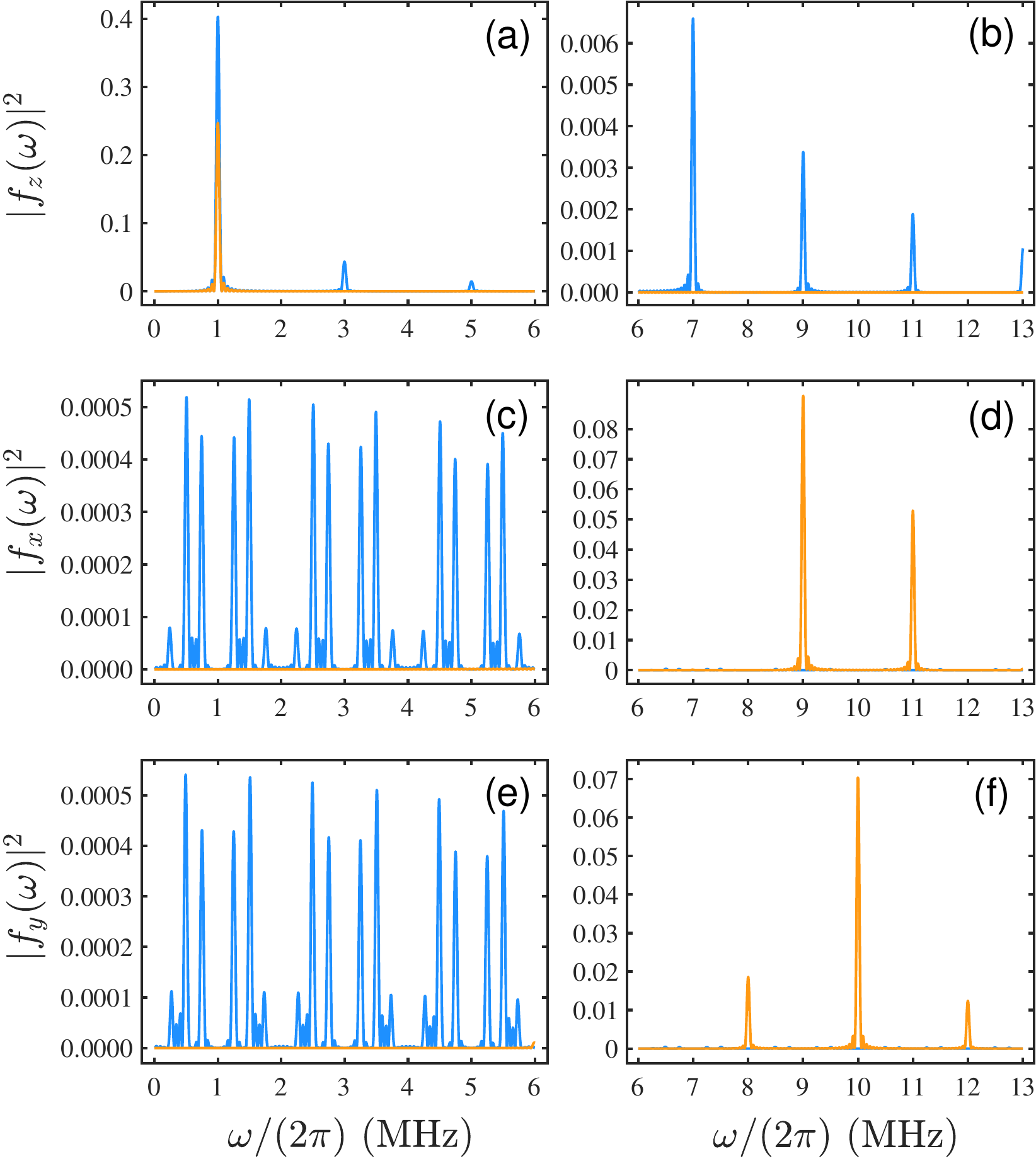}\caption{\label{fig:smFSpectrum} The power spectrum of $F_{\alpha}(t)$ ($\alpha=x,y,z$)
for the XY8 sequence in Fig.~\ref{fig:smXY8} (blue lines) and for
the geodesic evolution with $N=16$ in Fig.~\ref{fig:smGeo} (orange line).
$f_{\alpha}(\omega)=\frac{1}{T_{{\rm seq}}}\int_{0}^{T_{{\rm seq}}}F_{\alpha}(t)e^{i\omega t}dt$.
We choose the same total sequence time $T_{{\rm seq}}=16$ $\mu$s
for both methods. Both methods have the strongest spectral response for $f_{z}(\omega)$
at $\omega/(2\pi)=1/(2\tau)=\omega_{{\rm scan}}/(2\pi)=1$ MHz. For
the XY8 sequences, there are also resonances at $k/(2\tau)$ ($k=3,5,7,\ldots$)
due to higher harmonics. The nonzero pulse width ($\pi/\Omega$) also
results in complicated spurious peaks appeared in $f_{x}(\omega)$
and $f_{y}(\omega)$. For the method of geodesic evolution, only one
resonance response exists in a wide range of spectrum. For $f_{z}(\omega)$,
we have unambiguous frequency response at $\omega=\omega_{{\rm scan}}$,
and only tiny higher harmonics appear at $|\omega|\geq(N-1)|\omega_{{\rm scan}}|$, c.f. Fig.~1 in the main text. For $f_{x}(\omega)$
and $f_{y}(\omega)$, weak resonances appear at $(N\pm2)\omega_{{\rm scan}}$,
$(N\pm1)\omega_{{\rm scan}}$, and $N\omega_{{\rm scan}}$.}
\end{figure}

Here we provide a background on DD based quantum sensing. Consider
the qubit-enviroment Hamiltonian 
\begin{equation}
H_{{\rm int}}(t)=\frac{1}{2}\sigma_{z}B(t).
\end{equation}
A sequence of DD $\pi$ pulses is applied to the qubit, where every
$\pi$ pulse aims to flip the qubit and to induce a transformation
$\sigma_{z}\rightarrow-\sigma_{z}$ in the interaction picture with
respective to the DD control. Let $U_{{\rm DD}}(t)$ be the evolution
operator driven by the DD control. In the rotating frame of the DD
control, $H_{{\rm int}}(t)$ becomes 
\begin{equation}
\tilde{H}_{{\rm int}}(t)=\frac{1}{2}U_{{\rm DD}}^{\dagger}(t)\sigma_{z}U_{{\rm DD}}(t)B(t).
\end{equation}
For the idealized situation where the DD sequences use infinitely
sharp $\delta$-shape $\pi$ pulses (where the pulse duration $t_{p}$
equals 0), $\tilde{H}_{{\rm int}}(t)=\frac{1}{2}F(t)\sigma_{z}B(t)$
with $F(t)\in\{+1,-1\}$. In realistic experiments the $\pi$ pulses have
a nonzero duration $t_{p}$ because the control has a finite
Rabi frequency $\Omega$. In general, 
\begin{equation}
\tilde{H}_{{\rm int}}(t)=\frac{1}{2}\sum_{\alpha=x,y,z}F_{\alpha}(t)\sigma_{\alpha}B(t).
\end{equation}
When the $\pi$ pulses have a constant high $\Omega$, $t_{p}=\pi/\Omega$.
For the control Hamiltonian $H_{c}(t)=\Omega\frac{\sigma_{x}}{2}$,
we have during the first $\pi$ pulse $\tilde{H}_{{\rm int}}(t)=\frac{1}{2}\left[\sigma_{z}\cos(\Omega t)+\sigma_{y}\sin(\Omega t)\right]B(t)$.
A plot of the $F_{\alpha}(t)$ for a XY8 sequence is shown in Fig.
\ref{fig:smXY8}, and the corresponding Fourier transforms of $F_{\alpha}(t)$
are shown by the blue lines in Fig.~\ref{fig:smFSpectrum}. Under
the ideal situation that there is no control error and that the pulse
duration $t_{p}=0$, the modulation factor
\begin{equation}
F(t)=F_{z}(t)=\sum_{k=1}^{\infty}\frac{4\sin(k\frac{\pi}{2})}{k\pi}\cos(k\frac{\pi}{\tau}t),
\end{equation}
has harmonics of the frequency $\pi/\tau$ for an interpulse
spacing $\tau$ of equally spaced DD sequences. For $t_{p}>0$, $F_{x}(t)$ and $F_{y}(t)$ can have
non-zero values during the $\pi$ pulses and their Fourier transforms
have many peaks, which cause spurious signals in quantum sensing. 

\begin{figure}
\includegraphics[width=1\columnwidth]{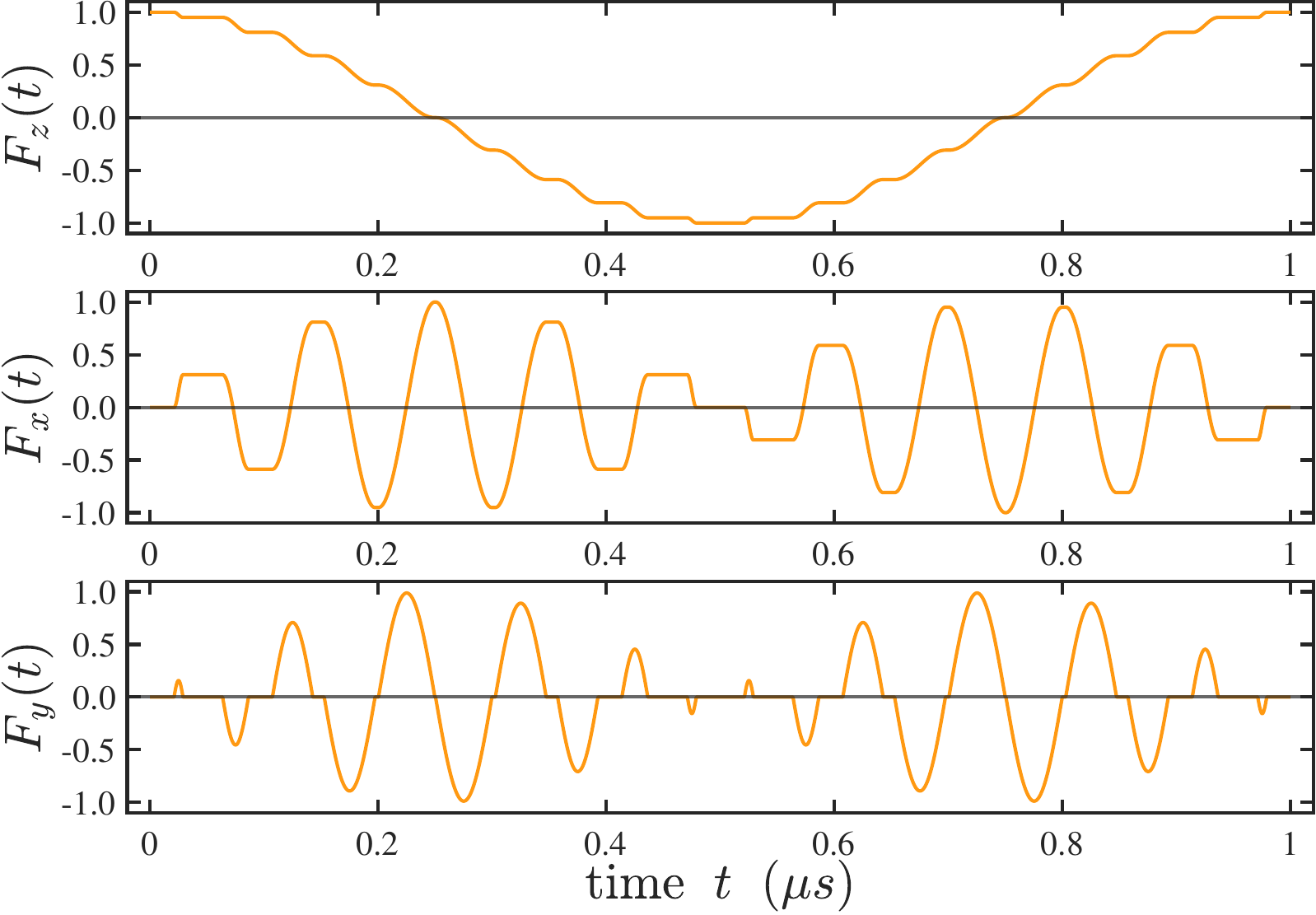}\caption{\label{fig:smGeo} The modulation functions $F_{\alpha}(t)$ induced
by the geodesic evolution using \wzy{$\text{GD}_{y}$}. The plots show a complete
period of $F_{\alpha}(t)$. The control field has a Rabi frequency
$\Omega=2\pi\times10$ MHz. The control parameters $N=20$ and $\omega_{{\rm scan}}/(2\pi)=1$
MHz.}
\end{figure}

\subsection{Quantum sensing by geometric evolution \label{subsec:Quantum-sensing-by}}

Here we provide details on the operator dynamics in the rotating
frame of the control $H_{c}(t)=\frac{E(t)}{2}(|+_{\phi}\rangle\langle+_{\phi}|-|-_{\phi}\rangle\langle-_{\phi}|)$
{[}Eq.~(1) in the main text{]} with $\phi=\phi(t)$ and
\begin{align}
|+_{\phi}\rangle= & \cos(\frac{\phi}{2})|+_{0}\rangle+\sin(\frac{\phi}{2})|-_{0}\rangle,\\
|-_{\phi}\rangle= & -\sin(\frac{\phi}{2})|+_{0}\rangle+\cos(\frac{\phi}{2})|-_{0}\rangle.
\end{align}
Without lost of generality, we assume that $\sigma_{z}=|1\rangle\langle1|-|0\rangle\langle0|$,
$\sigma_{x}=|1\rangle\langle0|+|0\rangle\langle1|$, and $\sigma_{x}=-i|1\rangle\langle0|+i|0\rangle\langle1|$
with $|+_{0}\rangle=|1\rangle$ and $|-_{0}\rangle=|0\rangle$. In
the rotating frame of the evolution $U_{c}(t)=\mathcal{T}e^{-i\int_{0}^{t}H_{c}(t^{\prime})dt^{\prime}}$,
we write 
\begin{equation}
U_{c}^{\dagger}\sigma_{z}U_{c}=\sum_{\alpha=x,y,z}F_{\alpha}(t)\sigma_{\alpha}.
\end{equation}
A plot of the $F_{\alpha}(t)$ for \wzy{$\text{GD}_{y}$} is shown in Fig.~\ref{fig:smGeo}.
The corresponding Fourier transforms are shown by the yellow lines in Fig.~\ref{fig:smFSpectrum}. One can see that the geodesic evolution
removes higher harmonics and spurious responses over a wide frequency
band. For the  Fourier transform of $F_{z}(t)$, we have unambiguous frequency response at $\omega=\omega_{{\rm scan}}$ given that $\phi(T_j)=\omega_{{\rm scan}}T_j$. 
Tiny higher harmonics only appear at the frequency $\omega>(N-1)\omega_{{\rm scan}}$. For the Fourier transforms of $F_{x}(t)$
and $F_{y}(t)$, weak resonances only appear at $\omega>(N-2)\omega_{{\rm scan}}$. Therefore, one can arbitary remove unwanted spectral respoances via a sufficently large $N$.

When the control $H_{c}(t)$ realizes the ideal adiabatic evolution
$U_{c}(\phi)=e^{-i\varphi_{+}(t)}|+_{\phi}\rangle\langle+_{0}|+e^{-i\varphi_{-}(t)}|-_{\phi}\rangle\langle-_{0}|$
\citep{sm:wang2016necessary,sm:xu2019breaking,sm:liu2022shortcuts},
we have 
\begin{align}
U_{c}^{\dagger}\sigma_{z}U_{c} & =\cos\phi\sigma_{z}-\sin\phi[e^{i\varphi(t)}|1\rangle\langle0|+{\rm h.c.}],\nonumber \\
U_{c}^{\dagger}\sigma_{x}U_{c} & =\sin\phi\sigma_{z}+\cos\phi[e^{i\varphi(t)}|1\rangle\langle0|+{\rm h.c.}],\\
U_{c}^{\dagger}\sigma_{y}U_{c} & =ie^{i\varphi(t)}|1\rangle\langle0|+{\rm h.c.}.\nonumber 
\end{align}
where the difference of dynamic phases $\varphi(t)=\varphi_{+}(t)-\varphi_{-}(t)=\int_{0}^{t}E(s)ds$
increases with time and $e^{i\varphi(t)}$ is a fast oscillating factor
with a zero mean. Neglecting terms involving $e^{i\varphi(t)}$ in
a way similar to the rotating wave approximation, we obtain Eq.~(5)
in the main text, i.e.,
\begin{align}
U_{c}^{\dagger}\sigma_{z}U_{c} & \approx\cos\phi\sigma_{z},\nonumber \\
U_{c}^{\dagger}\sigma_{x}U_{c} & \approx\sin\phi\sigma_{z},\\
U_{c}^{\dagger}\sigma_{y}U_{c} & \approx0,\nonumber 
\end{align}
and 
\begin{equation}
U_{c}^{\dagger}(\sigma_{z}+i\sigma_{x})U_{c}=\sigma_{z}e^{i\phi}.\label{eq:sigmaEIPhi}
\end{equation}

The operator dynamics can also be analyzed as follows. The control
$H_{c}(t)$ with $\phi=\phi_{j}$ is used to realize the $j$th $\pi$
pulse which has the propagator $U_{j}=\exp\left[-i\frac{\pi}{2}\left(\cos\phi_{j}\sigma_{z}+\sin\phi_{j}\sigma_{x}\right)\right]$.
After $n$ such $\pi$ pulses, we have
\begin{equation}
U_{c}=U_{n}U_{n-1}\cdots U_{2}U_{1}.
\end{equation}
This gives the transformation after $n$ $\pi$ pulses, $U_{c}^{\dagger}\sigma_{z}U_{c}=\sigma_{z}\cos\Theta_{n}+\sigma_{x}\sin\Theta_{n}$,
where $\Theta_{n}=2\sum_{j=1}^{n}(-1)^{j+1}\phi_{j}$. Using the relation
$\phi_{j}=\omega_{{\rm scan}}T_{{\rm scan}}\frac{2j-1}{2N}$, we obtain
$\Theta_{n}=(-1)^{n+1}\frac{n}{N}\omega_{{\rm scan}}T_{{\rm scan}}$,
where the factor $(-1)^{n}$ is caused by the $\pi$ phase shifts
of the dynamic phases. Neglecting the fast oscillating term $\sigma_{x}\sin\Theta_{n}$
in a way similar to the rotating wave approximation, we obtain after
$n$ $\pi$ pulses $U_{c}^{\dagger}\sigma_{z}U_{c}\approx\sigma_{z}\cos\phi_{n}$.
Since $\phi_j=\phi(T_j)=\omega_{{\rm scan}} T_j$, we have $U_{c}^{\dagger}\sigma_{z}U_{c}\approx\sigma_{z}\cos\phi(t)$. 

\subsection{Sensing of single nuclear spins by \wzy{$\text{GD}_{y}$}}

Here we provide details of quantum sensing by geometric evolution
for the case of sensing of single nuclear spins by a single NV center.
The Hamiltonian of a NV electron spin and its surrounding nuclear
spins reads

\begin{equation}
H=-\sum_{j}\gamma_{j}\vec{I}_{j}\cdot(B_{z}\hat{z})+|1\rangle\langle1|\sum_{j}\vec{A}_{j}\cdot\vec{I}_{j}+H_{c}(t),\label{eq:H_nuclear}
\end{equation}
where $\vec{I}_{j}$ is the spin-$\frac{1}{2}$ operator for the $j$th
nuclear spin, $\gamma_{j}$ is the gyromagnetic ratio, and $\vec{A}_{j}=(A_{j}^{x},A_{j}^{y},A_{j}^{z})$
describes the hyperfine interaction at the spin location. $B_{z}\hat{z}$
is a magnetic field with an amplitude $B_{z}$ and a direction denoted
by the unit vector $\hat{z}$. $H_{c}(t)=\Delta(t)\frac{\sigma_{z}}{2}+\Omega(t)\frac{\sigma_{x}}{2}$
as in the main text for \wzy{$\text{GD}_{y}$}. We has neglected the week nuclear-nuclear coupling in the Hamiltonian
because it is much smaller than the electron-nuclear coupling. 

The nuclear spins have the effective precession frequency $\vec{\nu}_{j}\equiv\nu_{j}\hat{\nu}_{j}=\gamma_{j}B_{z}\hat{z}-\frac{1}{2}\vec{A}_{j}$,
where $\hat{\nu}_{j}$ is a unit vector. In the rotating frame with
respect to $-\sum_{j}\vec{\nu}_{j}\cdot\vec{I}_{j}$, the Hamiltonian
becomes
\begin{equation}
\tilde{H}(t)=H_{{\rm int}}(t)+H_{c}(t),
\end{equation}
\begin{equation}
H_{{\rm int}}(t)=\frac{\sigma_{z}}{2}B(t),
\end{equation}
where 
\begin{equation}
B(t)=\sum_{j}\frac{1}{2}a_{j}^{x}(I_{j}^{+}e^{-i\nu_{j}t}+\text{h.c.})+\frac{1}{2}a_{j}^{z}I_{j}^{z},
\end{equation}
$a_{j}^{z}=\vec{A}_{j}\cdot\hat{\nu}_{j}$, and $a_{j}^{x}=|\vec{A}_{j}-a_{j}^{z}\hat{\nu}_{j}|$.
Defining $\hat{z}_{j}\equiv\hat{\nu}_{j}$, $\hat{x}_{j}\equiv(\vec{A}_{j}-a_{j}^{z}\hat{\nu}_{j})/a_{j}^{x}$,
and $\hat{y}_{j}\equiv\hat{z}_{j}\times\hat{x}_{j}$, we write $I_{j}^{\mu}=\vec{I}_{j}\cdot\hat{\mu}_{j}$
with $\mu\in\{x,y,z\}$ and $I_{j}^{\pm}=I_{j}^{x}\pm iI_{j}^{y}$.
In the rotating frame with respect to $H_{c}(t)$, the Pauli operator $\sigma_{z}$
becomes
\begin{equation}
U_{c}^{\dagger}\sigma_{z}U_{c}\equiv\sum_{\alpha=x,y,z}F_{\alpha}(t)\sigma_{\alpha},
\end{equation}
where $F_{\alpha}(t)$ are oscillating factors. See Figs.~\ref{fig:smGeo}
and \ref{fig:smXY8} for exemplified plots for \wzy{$\text{GD}_{y}$} and
XY8 sequences, respectively. For the case of \wzy{$\text{GD}_{y}$}, 
\begin{equation}
U_{c}^{\dagger}\sigma_{z}U_{c}\approx F(t)\sigma_{z}
\end{equation}
in the relevant frequency range. With a sufficiently large $N$, 
\begin{equation}
F(t)=F_{z}(t)\approx\cos(\omega_{{\rm scan}}t).
\end{equation}

Therefore, when $\omega_{{\rm scan}}=\nu_{n}$ for the $n$th spin
and assuming that $\omega_{{\rm scan}}\gg a_{j}^{\alpha}$ and $a_{j}^{\alpha}$
are much smaller than the differences between $\nu_{n}$ and other
$\nu_{j}$, we obtain 
\begin{equation}
\tilde{H}_{{\rm int}}\approx\frac{1}{2}a_{n}^{x}\sigma_{z}I_{n}^{x},
\end{equation}
where the qubit sensor couples to only one nucler spin. 

\subsection{Heterodyne sensing by \wzy{$\text{GD}_{z}$} }

We consider a single-qubit sensor exposed to a high-frequency signal
field $\vec{B}(t)=(B_{x},B_{y},B_{z})$ and a dephasing noise $\delta_{t}$,
with the Hamiltonian 
\begin{align}
H_{0} & =\left(\omega_{q}+\delta_{t}+B_{z}\right)\frac{\sigma_{z}}{2}+B_{x}\frac{\sigma_{x}}{2}+B_{y}\frac{\sigma_{y}}{2},\\
 & =\left(\omega_{q}+\delta_{t}+B_{z}\right)\frac{\sigma_{z}}{2}+(\frac{1}{2}B_{\perp}\sigma_{+}+{\rm h.c.}),
\end{align}
where $B_{\perp}\equiv B_{x}-iB_{y}=\sum_{j}b_{j}e^{i\alpha_{j}}\cos(\nu_{j}t+\theta_{j})$
and $\sigma_{+}=\frac{1}{2}(\sigma_{x}+i\sigma_{y})$. Adding a control
$H_{{\rm ctr}}=\Omega(t)\cos(\omega_{{\rm ctr}}t+\phi)\sigma_{x}$
with a detuning $\Delta_{{\rm scan}}=\omega_{{\rm ctr}}-\omega_{q}$,
the total Hamiltonian is $H=H_{0}+H_{{\rm ctr}}$. In the rotating
frame with respect to $\frac{1}{2}\omega_{{\rm ctr}}\sigma_{z}$,
the Hamiltonian becomes 
\begin{align}
H^{\prime}= & \left(\delta_{t}+B_{z}-\Delta_{{\rm scan}}\right)\frac{\sigma_{z}}{2}\nonumber \\
 & +\left[\frac{1}{2}B_{\perp}(t)\sigma_{+}e^{i\omega_{{\rm ctr}}t}+{\rm h.c.}\right]+\Omega(t)\frac{\sigma_{\phi}}{2},
\end{align}
where 
\begin{equation}
\sigma_{\phi}\equiv\sigma_{x}\cos\phi+\sigma_{y}\cos\phi.
\end{equation}
The control part $H_{c}=\Omega(t)\frac{\sigma_{\phi}}{2}$ realizes
a sequence of $\pi$ pulses for accelerated quantum adiabatic evolution
along the geodesic and prolongs the coherence time of the qubit by
suppressing the nosie in a way similar to DD \citep{sm:xu2019breaking}.
On the other hand, $B_{z}$ is a fast oscillating field at a frequency
much larger than the Rabi frequency $\Omega(t)$. Therefore, the effect
of the term $\left(\delta_{t}+B_{z}-\Delta_{{\rm scan}}\right)\frac{\sigma_{z}}{2}$
can be neglected. Similar to Eq.~(\ref{eq:sigmaEIPhi}) in Sec. \ref{subsec:Quantum-sensing-by},
in the rotating frame of $U_{c}(t)=\mathcal{T}e^{-i\int_{0}^{t}H_{c}(t^{\prime})dt^{\prime}}$,
$U_{c}^{\dagger}\sigma_{x}U_{c}\approx\cos\phi\sigma_{x}$ and $U_{c}^{\dagger}\sigma_{y}U_{c}\approx\sin\phi\sigma_{x}$,
which implies 
\begin{equation}
U_{c}^{\dagger}\sigma_{+}U_{c}\approx\frac{1}{2}\sigma_{x}e^{i\phi}\approx\frac{1}{2}\sigma_{x}e^{i\omega_{{\rm scan}}t}.
\end{equation}
Therefore, in the rotating frame of $U_{c}(t)$, we obtain the effective
Hamiltonian 
\begin{align}
\tilde{H}_{{\rm int}}(t) & \approx U_{c}^{\dagger}\left[\frac{1}{2}B_{\perp}(t)\sigma_{+}e^{i\omega_{{\rm ctr}}t}+{\rm h.c.}\right]U_{c}\\
 & \approx \frac{1}{4}B_{\perp}(t)\sigma_{x}e^{i(\omega_{{\rm ctr}}+\omega_{{\rm scan}})t}+{\rm h.c.},\nonumber\\
 & \approx\frac{1}{8}\sum_{j}b_{j}e^{i(\alpha_{j}-\theta_{j})+i(\omega_{{\rm ctr}}+\omega_{{\rm scan}}-\nu_{j})t}\sigma_{x}+{\rm h.c.},\nonumber \\
 & =\sum_{j}\frac{b_{j}}{4}\cos\left[(\omega_{{\rm ctr}}+\omega_{{\rm scan}}-\nu_{j})t+(\alpha_{j}-\theta_{j})\right]\sigma_{x}.\nonumber 
\end{align}
When the frequency of control matches a signal frequency $\nu_{n}$,
i.e., 
\begin{equation}
\omega_{{\rm ctr}}+\omega_{{\rm scan}}=\nu_{n},
\end{equation}
and the signal is weak $|b_{j}/(\nu_{j}-\nu_{n})|\ll1$, the qubit
sensor extracts the signal at this frequency with the Hamiltonian
\begin{equation}
\tilde{H}_{{\rm int}}(t)\approx\frac{1}{2}b_{n}\cos(\alpha_{n}-\theta_{n})\frac{\sigma_{x}}{2},
\end{equation}
by applying the rotating wave approximation. 

For a state $\rho$ after the control of a duration $T$, the population
remaining at the initial state is $P_{|\psi_{0}\rangle\langle\psi_{0}|}\equiv{\rm Tr}(\rho(T)|\psi_{0}\rangle\langle\psi_{0}|)$.
For the resonant case $\omega_{{\rm ctr}}+\omega_{{\rm scan}}=\nu_{n}$,
$P_{|\psi_{0}\rangle\langle\psi_{0}|}=\cos^{2}\left[\frac{1}{4}b_{n}\cos(\theta_{n}-\alpha_{n})T\right]$,
while for the off-resonant case, $P_{|\psi_{0}\rangle\langle\psi_{0}|}\approx1$. 

\section{Simulation details}

\subsection{AC field sensing \label{subsec:AC-field-sensing}}

The AC field used for Figs.~2 (a)-(b) of the main text is $B(t)=\sum_{j}b_{j}\cos(\nu_{j}t+\theta_{j})$,
where $\{\nu_{j}\}=2\pi\times\{500,1500.05,2499.875\}$ kHz, $\{b_{j}\}=2\pi\times\{5,10,10\}$
kHz, and $\theta_{j}=0$. 

Without control errors and dephasing noise, the Hamiltonian for the
simulation reads $H=B(t)\frac{\sigma_{z}}{2}+H_{c}(t)$. For the
XY8 sequences, $H_{c}(t)=\frac{1}{2}\Omega(t)\sigma_{\mu}$ when applying 
$\mu$ ($\mu=x,y$) pulses  (see Fig.
\ref{fig:smXY8} for the pulse sequence). For \wzy{$\text{GD}_{y}$}, $H_{c}(t)=\Delta(t)\frac{\sigma_{z}}{2}+\Omega(t)\frac{\sigma_{x}}{2}$.
In the simulations of both protocols, we choose $\Omega(t)=\Omega=2\pi\times10$
MHz for the Rabi frequency during the $\pi$ pulses when there is
no control errors. In \wzy{$\text{GD}_{y}$}, we use the detuning of the control
$\Delta=\Omega\cot\phi_{j}$ for the $j$th pulse according to the
ideal value of $\Omega$. For the sensing using XY8 sequences, we
use the resonant frequency $1/(2\tau)$ at the 1st harmonic to detect
the signal, and we repeat the XY8 unit 16 times implying the whole
XY8 sequence has $N^{\prime}=128$ $\pi$ pulses and a total duration
$T_{{\rm seq}}=128\tau$. To compare the results for the same sequence
duration $T_{{\rm seq}}$, we apply $M=64$ cycles of the cyclic
evolution when we use the \wzy{$\text{GD}_{y}$} (with $N=25$). In both protocols
the qubit is initially prepared in the superposition state $|+\rangle=(|0\rangle+|1\rangle)/\sqrt{2}$
and we measure the population remaining at the initial state $|+\rangle$
as the signal.

For the results with dephasing noise and control errors, we add
a noise term $\delta B(t)$ to $B(t)$ and an amplitude flutuation
to the Rabi frequency $\Omega(t)$ in both protocols. $\delta B(t)$
is modeled by a stochastic Ornstein-Uhlenbeck (OU) process \citep{sm:gardiner2004handbook}
with a correlation time $\tau_{B}=4$ ms and a standard deviation
$\delta_{B}=2\pi\times0.1$ MHz \citep{sm:chu2021precise}, which
gives a decoherence time $T_{2}^{*}\approx2$ $\mu$s, i.e., a time
when the phase coherence decays to $1/e$ of its original value in
the absence of quantum control. In the modeling of the control fluctuation,
we change the value of $\Omega(t)$ to $[1+\eta(t)](1+\epsilon)\Omega(t)$,
where the static bias $\epsilon=20\%$ and the time-dependent fluctuation
$\eta(t)$ is modeled by a stochastic OU process with a correlation
time of $1$ ms and a standard deviation $2.4\times10^{-3}$ \citep{sm:cai2012robust}.
The results are the average of 6500 Monte Carlo simulations. The
initial state is $|\psi_{0}\rangle=(|0\rangle+|1\rangle)/\sqrt{2}$,
and we measure the population remaining at the initial state as the
signal. 

\subsection{Arbitrary high-resolution spectroscopy}

Combining the sensing sequences with the method of synchronized readout
\citep{sm:boss2017quantum,sm:schmitt2017submillihertz} can improve
the spectral resolution. In the simulations for Figs.~2 (c) and (d)
of the main text, we use the same AC field {[}$B(t)=\sum_{j}b_{j}\cos(\nu_{j}t+\theta_{j})$
with $\{\nu_{j}\}=2\pi\times\{500,1500.05,2499.875\}$ kHz, $\{b_{j}\}=2\pi\times\{5,10,10\}$
kHz, $\theta_{j}=0${]} and sequence parameters ($\Omega=2\pi\times10$
MHz, $N^{\prime}=128$, $M=64$, and $N=25$) as in Sec. \ref{subsec:AC-field-sensing}.

We repeatedly apply the same sequence (each has the duration
$T_{{\rm seq}}=128\tau$) for $10^{4}$ times to obtain the synchronized
readout signals [see the lower panel of Fig. 1(a) in the main text]. At the end of each sequence we make a projection measurement
and record the measured values. There is a delay of $t_{r}$ for readout
and initialization between the end of a sequence and the starting
time of its subsequent sequence. The readout process at the end of the $n$th sequence
yields at the moment $t$ a random photon number $y_{n}(t)={\rm Pois}[C(1-r{\rm Bn}[p_{n}]]$,
where Pois denotes a Poisson process and Bn is a Bernoulli process
that takes the value $1$ with a probability of $p_{n}$ and the value
0 with probability $1-p_{n}$ \citep{sm:boss2017quantum}. We initialize
the qubit sensor at the state $|\psi_{0}\rangle=(|0\rangle+|1\rangle)/\sqrt{2}$
before each sequence. The probability $p_{n}$ is the population signal
of finding the qubit in the state $|+_{y}\rangle=(|0\rangle+i|1\rangle)/\sqrt{2}$
at the end of the sequence. We choose $r=0.35$, $t_{r}=2.32$ $\mu$s,
and $C\approx0.105n$ with $n=10$ according to the parameters in
Ref. \citep{sm:boss2017quantum}. A discrete Fourier transform $\tilde{y}$
of the photon numbers $\{y_{n}(t)\}$ results in the power spectrum
$|\tilde{y}|/\mathcal{N}$ shown in Figs.~2 (c) and (d). Here $\mathcal{N}$ is a fixed scaling factor for all the protocols.

\subsection{Sensing of single nuclear spins}
In Figs.~2 (e) and (f) of the main text for the sensing of a proton
spin, we perform the simulations by using the Hamiltonian Eq.~(\ref{eq:H_nuclear}).
The control $H_{c}(t)=\Delta(t)\frac{\sigma_{z}}{2}+\Omega(t)\frac{\sigma_{x}}{2}$
as in the main text for \wzy{$\text{GD}_{y}$}. For the XY8 sequences,
$H_{c}(t)=\frac{1}{2}\Omega(t)\sigma_{\mu}$ ($\mu\in\{x,y\}$). 
The parameters for the proton spin are $\gamma_{^{1}\text{H}}/2\pi\approx4.2576$
kHz/G, $\sqrt{(A_{\text{H}}^{x})^{2}+(A_{\text{H}}^{y})^{2}}/2\pi=0.5$
kHz, and $A_{\text{H}}^{z}/2\pi=0.5$ kHz. We assume that one $^{13}$C
nucleus (with a natural abundance $1.1$\% and $\gamma_{\text{C}}\approx1.0705$
kHz/G) in the diamond is coupled to the NV with $\sqrt{(A_{\text{C}}^{x})^{2}+(A_{\text{C}}^{y})^{2}}/2\pi=100$
kHz and $A_{\text{C}}^{z}/2\pi=28$ kHz. The magnetic field $B_{z}=100$
G. The Rabi frequency for the
$\pi$ pulse is $\Omega=2\pi\times10$ MHz. The XY8 sequence is repeated
50 times for a total number of $N^{\prime}=400$ $\pi$ pulses. We
change $B(t)\rightarrow B(t)+\Delta_{{\rm error}}$ to include detuning
error. We set $\Delta_{{\rm error}}=2\pi\times2$ MHz as it is the
typical error due to unpolarized nitrogen spin of the NV center. An
amplitude error $\epsilon$ will change $\Omega(t)$ to $(1+\epsilon)\Omega(t)$,
where we set $\epsilon=30 \%$. The initial state of the nuclear spins
are assumed to be in a room-temperature thermal state. We measure
the population of the NV qubit sensor remaining at its initial state
$|\psi_{0}\rangle=(|0\rangle+|1\rangle)/\sqrt{2}$ as the signal. 

\subsection{High-frequency AC field sensing }

In Fig.~3 of the main text, the signal field $\vec{B}(t)=(B_{x},B_{y},B_{z})$
has the transverse component $B_{\perp}\equiv B_{x}-iB_{y}=\sum_{j}b_{j}e^{i\alpha_{j}}\cos(\nu_{j}t+\theta_{j})$.
In the simulation, $\{(\nu_{j}-\omega_{q})/2\pi\}=\{-84,-68,-56,-50,-42,-2,72,90\}$
kHz, with the corresponding $\{b_{j}/2\pi\}=\{5.3,4.6,2.7,4.2,3.4,4.5,2.2,3.1\}$
kHz. The angles $\alpha_{j}$ and phases $\theta_{j}$ are random
real numbers. The control field has a Rabi frequency $\Omega=2\pi\times50$
MHz. For the results with dephasing noise $\delta_{t}$, in each run
of the simulation we model $\delta_{t}$ by a OU process with a
correlation time $\tau_{B}=4$ ms and a standard deviation $\delta_{B}=2\pi\times0.1$
MHz for a decoherence time $T_{2}^{*}\approx2$ $\mu$s. The results
are the average of $10^{4}$ Monte Carlo simulations. 

For the heterodyne spectroscopy using CPMG sequences, the
initial state is the superposition state $|\psi_{0}\rangle=(|0\rangle+|1\rangle)/\sqrt{2}$
and the population remained at $|\psi_{0}\rangle$ is the measured
signal. 

For the heterodyne spectroscopy using \wzy{$\text{GD}_{z}$}, the initial
state is $|\psi_{0}\rangle=|0\rangle$. For the case of NV center,
$|0\rangle$ is the state after optical illumination and we do not
need to use the initial $\pi/2$ pulse to prepare the qubit state.
We measure the population remained at $|\psi_{0}\rangle$ as the signal.
There are 16 cycles of geodesic evolution where for each cycle there
are $N=10$ $\pi$ pulses. 

\subsection{Robustness and Combination with composite pulses }

In Fig.~4 of the main text we demonstrate the control robustness in the absence of signals. A Rabi frequency $\Omega=2\pi\times50$
MHz is used for the simulations. For DD sequences (CPMG and XY8 sequences)
we choose $N^{\prime}=40$ pulses for each sequence and the pulse interval
$\tau$ satisfies $1/(2\tau)=1$ MHz. Therefore each DD sequence has
a time length $T_{{\rm seq}}=2\tau\times N^{\prime}/2=40$ $\mu$s.
For the protocols of geodesic driving, we use 
$\omega_{{\rm scan}}/(2\pi)=1$ MHz for the same sensing frequency as DD, $\Delta_{{\rm scan}}=0$, and the same $T_{{\rm seq}}=40$ $\mu$s.
To investigate the robustness of the protocols, we change the Rabi
frequency to $(1+\epsilon)\Omega$ by adding a relative fluctuation,
and add a detuning error $\Delta$ (i.e., adding $\Delta\frac{\sigma_{z}}{2}$
to the qubit Hamiltonian). We denote $U(\frac{\Delta}{\Omega},\epsilon)$
the actual evolution realized by the sequences in the presence of
control errors. In the absence of control errors, the ideal evolution
after the sequence is an identity operator, $U_{{\rm ideal}}=I$.
The control fidelity is calculated in the standard way as \citep{sm:wang2008an}
\begin{equation}
{\rm fidelity}={\rm Tr}\left[U\left(\frac{\Delta}{\Omega},\epsilon\right)U_{{\rm ideal}}^{\dagger}\right]/{\rm Tr}I.
\end{equation}


\begin{thebibliography}{99}
\bibitem{preskill2018quantum} J. Preskill, Quantum Computing in the
NISQ era and beyond, Quantum \textbf{2}, 79 (2018).

\bibitem{paladino2014} E. Paladino, Y. M. Galperin, G. Falci, and
B. L. Altshuler, 1/f noise: Implications for solid-state quantum information,
Rev. Mod. Phys. \textbf{86}, 361 (2014).

\bibitem{degen2017quantum} C. L. Degen, F. Reinhard, and P. Cappellaro,
Quantum sensing, Rev. Mod. Phys. \textbf{89}, 035002 (2017).

\bibitem{barry2020sensitivity} J. F. Barry, J. M. Schloss, E. Bauch,
M. J. Turner, C. A. Hart, L. M. Pham, and R. L. Walsworth, Sensitivity
optimization for NV-diamond magnetometry, Rev. Mod. Phys. \textbf{92}, 015004
(2020). 

\bibitem{viola1999dynamical} L. Viola, E. Knill, and S. Lloyd, Dynamical
decoupling of open quantum systems, Phys. Rev. Lett. \textbf{82}, 2417 (1999).

\bibitem{yang2011preserving} W. Yang, Z.-Y. Wang, and R.-B. Liu,
Preserving qubit coherence by dynamical decoupling, Front. Phys. \textbf{6},
2 (2011).

\bibitem{javaloy2021dynamical} C. Munuera-Javaloy, R. Puebla, and
J. Casanova, Dynamical decoupling methods in nanoscale NMR, Europhys.
Lett. \textbf{134}, 30001 (2021).

\bibitem{doherty2013the}M. W. Doherty, N. B. Manson, P. Delaney, F. Jelezko, J. Wrachtrup, and L. C.Hollenberg, The Nitrogen-Vacancy Colour Centre in Diamond, Phys. Rep. \textbf{528}, 1 (2013).

\bibitem{WuJPW16} Y. Wu, F. Jelezko, M.B. Plenio, and T. Weil, Diamond Quantum Devices in Biology, Angewandte Chemie – International 
Edition {\bf 55}, 6586 – 6598 (2016).

\bibitem{weber2010quantum} J. R. Weber, W. F. Koehl, J. B. Varley,
A. Janotti, B. B. Buckley, C. G. Van de Walle, D. D. Awschalom, Quantum
computing with defects. Proc. Natl. Acad. Sci. U.S.A. \textbf{107}, 8513--8518
(2010).

%% === Nano NMR



\bibitem{kolkowitz2012sensing} S. Kolkowitz, Q. P. Unterreithmeier,
S. D. Bennett, and M. D. Lukin, Sensing Distant Nuclear Spins with
a Single Electron Spin, Phys. Rev. Lett. \textbf{109}, 137601 (2012). 

\bibitem{zhao2012sensing} N. Zhao, J. Honert, B. Schmid, M. Klas,
J. Isoya, M. Markham, D. Twitchen, F. Jelezko, R. B. Liu, H. Fedder,
and J. Wrachtrup, Sensing single remote nuclear spins, Nat. Nanotechnol.
\textbf{7}, 657 (2012).

\bibitem{staudacher2013nuclear} T. Staudacher, F. Shi, S. Pezzagna,
J. Meijer, J. Du, C. A. Meriles, F. Reinhard, and J. Wrachtrup, Nuclear
Magnetic Resonance Spectroscopy on a (5-Nanometer)3 Sample Volume,
Science \textbf{339}, 561 (2013).

\bibitem{muller2014nuclear} C. Müller, X. Kong, J.-M. Cai, K. Melentijević, A. Stacey, 
M. Markham, D. Twitchen, J. Isoya, S. Pezzagna, J. Meijer, J. Du, 
M. B. Plenio, B. Naydenov, L. P. McGuinness, and F. Jelezko,
Nuclear magnetic resonance spectroscopy with single spin sensitivity,
Nat. Commun. \textbf{5}, 4703 (2014). 

\bibitem{rugar2015} D. Rugar, H. J. Mamin, M. H. Sherwood, M. Kim,
C. T. Rettner, K. Ohno, and D. D. Awschalom, Proton magnetic resonance
imaging using a nitrogen-vacancy spin sensor, Nat. Nanotechnol. \textbf{10},
120 (2015).

\bibitem{deVience2015nanoscale} S. J. DeVience, L. M. Pham, I. Lovchinsky,
A. O. Sushkov, N. Bar-Gill, C. Belthangady, F. Casola, M. Corbett,
H. Zhang, M. Lukin, H. Park, A. Yacoby, and R. L. Walsworth, Nanoscale
NMR spectroscopy and imaging of multiple nuclear species, Nat. Nanotechnol.
\textbf{10}, 129 (2015).

\bibitem{glenn2018high} D. R. Glenn, D. B. Bucher, J. Lee, M. D.
Lukin, H. Park, R. L. Walsworth, High-resolution magnetic resonance
spectroscopy using a solid-state spin sensor. Nature \textbf{555}, 351--354
(2018).

\bibitem{lang2017enhanced} E. Lang, J. Casanova, Z.-Y. Wang, M. B.
Plenio, and T. S. Monteiro, Enhanced Resolution in Nanoscale NMR via
Quantum Sensing with Pulses of Finite Duration, Phys. Rev. Applied
\textbf{7}, 054009 (2017). 

\bibitem{pfender2019high} M. Pfender, P. Wang, H. Sumiya, S. Onoda,
W. Yang, D. B. Rao Dasari, P. Neumann, X.-Y. Pan, J. Isoya, R.-B.
Liu, and J. Wrachtrup, High-resolution spectroscopy of single nuclear
spins via sequential weak measurements, Nat. Commun. \textbf{10}, 594 (2019).

\bibitem{bucher2019quantum} D. B. Bucher, D. P. L. Aude Craik, M.
P. Backlund, M. J. Turner, O. Ben Dor, D. R. Glenn, R. L. Walsworth,
Quantum diamond spectrometer for nanoscale NMR and ESR spectroscopy.
Nat. Protoc. \textbf{14}, 2707--2747 (2019).

\bibitem{casanova2019modulated} J. Casanova, E. Torrontegui, M. B. Plenio, J. J. García-Ripoll, and E. Solano, Modulated
Continuous Wave Control for Energy-Efficient Electron-Nuclear Spin
Coupling, Phys. Rev. Lett. \textbf{122} 010407 (2019).

\bibitem{aharon2019quantum} N. Aharon, I. Schwartz, and A. Retzker,
Quantum Control and Sensing of Nuclear Spins by Electron Spins under
Power Limitations, Phys. Rev. Lett. \textbf{122}, 120403 (2019).

\bibitem{javaloy2020robust} C. Munuera-Javaloy, Y. Ban, X. Chen,
and J. Casanova, Robust Detection of High-Frequency Signals at the
Nanoscale, Phys. Rev. Appl. \textbf{14}, 054054 (2020).

\bibitem{meinel2022quantum} J. Meinel, V. Vorobyov, P. Wang, B. Yavkin,
M. Pfender, H. Sumiya, S. Onoda, J. Isoya, R.-B. Liu, and J. Wrachtrup,
Quantum nonlinear spectroscopy of single nuclear spins, Nat. Commun.
\textbf{13}, 5318 (2022).

\bibitem{wang2023using} P. Wang, W. Yang, and R. Liu, Using Weak
Measurements to Synthesize Projective Measurement of Nonconserved
Observables of Weakly Coupled Nuclear Spins, Phys. Rev. Applied \textbf{19},
054037 (2023).

\bibitem{javaloy2023high} C. Munuera-Javaloy, A. Tobalina, and J.
Casanova, High-Resolution NMR Spectroscopy at Large Fields with Nitrogen
Vacancy Centers, Phys. Rev. Lett. \textbf{130}, 133603 (2023). 

%% == Spin labels 

\bibitem{shi2015single} F. Shi, Q. Zhang, P. Wang, H. Sun, J. Wang,
X. Rong, M. Chen, C. Ju, F. Reinhard, H. Chen, J. Wrachtrup, J. Wang,
and J. Du, Single-protein spin resonance spectroscopy under ambient
conditions, Science \textbf{347}, 1135 (2015).

\bibitem{lovchinsky2016nuclear} I. Lovchinsky, A. O. Sushkov, E. Urbach, N. P. de Leon, S. 
Choi, K. De Greve, R. Evans, R. Gertner, E. Bersin, C. Müller, L. 
McGuinness, F. Jelezko, R. L. Walsworth, H. Park, M. D. Lukin, Nuclear magnetic resonance detection and spectroscopy
of single proteins using quantum logic. Science \textbf{351}, 836 (2016)

\bibitem{javaloy2022detection} C. Munuera-Javaloy, R. Puebla, B. D. Anjou, M. B. Plenio, and J. Casanova,
Detection of molecular transitions with nitrogen-vacancy centers and
electron-spin labels, npj Quantum Inf. \textbf{8}, 140 (2022).

%% == Spin Clusters

\bibitem{zhao2011atomic} N. Zhao, J.-L. Hu, S.-W. Ho, J. T. K. Wan,
and R. B. Liu, Atomic-scale magnetometry of distant nuclear spin clusters
via nitrogen-vacancy spin in diamond, Nat. Nanotechnol. \textbf{6}, 242 (2011).

\bibitem{shi2014sensing} F. Shi, X. Kong, P. Wang, F. Kong, N. Zhao, R. B. Liu, and J. Du, Sensing and Atomic-Scale Structure
Analysis of Single Nuclear-Spin Clusters in Diamond, Nat. Phys. \textbf{10},
21 (2014).

\bibitem{lang2015dynamical} J. E. Lang, R. B. Liu, and T. S. Monteiro,
Dynamical Decoupling-Based Quantum Sensing: Floquet Spectroscopy,
Phys. Rev. X \textbf{5}, 041016 (2015).

\bibitem{wang2016positioning} Z.-Y. Wang, J. F. Haase, J. Casanova,
and M. B. Plenio, Positioning nuclear spins in interacting clusters
for quantum technologies and bioimaging, Phys. Rev. B \textbf{93}, 174104 (2016).

\bibitem{sasaki2018determination} K. Sasaki, K. M. Itoh, and E. Abe,
Determination of the position of a single nuclear spin from free nuclear
precessions detected by a solid-state quantum sensor, Phys. Rev. B
\textbf{98}, 121405(R) (2018).

\bibitem{zopes2018three}J. Zopes, K. S. Cujia, K. Sasaki, J. M. Boss,
K. M. Itoh, and C. L. Degen, Three-dimensional localization spectroscopy
of individual nuclear spins with sub-Angstrom resolution, Nat. Commun.
\textbf{9}, 4678 (2018). 

\bibitem{zopes2018threePRL} J. Zopes, K. Herb, K. S. Cujia, and C.
L. Degen, Three Dimensional Nuclear Spin Positioning Using Coherent
Radio-Frequency Control, Phys. Rev. Lett. \textbf{121}, 170801 (2018).

\bibitem{abobeih2019atomic} M. H. Abobeih, J. Randall, C. E. Bradley,
H. P. Bartling, M. A. Bakker, M. J. Degen, M. Markham, D. J. Twitchen,
and T. H. Taminiau, Atomic-scale imaging of a 27-nuclear-spin cluster
using a quantum sensor, Nature \textbf{576}, 411 (2019)

\bibitem{cujia2022parallel} K. S. Cujia, K. Herb, J. Zopes, J. M.
Abendroth, and C. L. Degen, Parallel detection and spatial mapping
of large nuclear spin clusters, Nat. Commun. \textbf{13} 1260 (2022).

%% == AC field

\bibitem{schmitt2017submillihertz} S. Schmitt, T. Gefen, F. M. Stürner, T. Unden, G. Wolff, C. 
Müller, J. Scheuer, B. Naydenov, M. Markham, S. Pezzagna, J. 
Meijer, I. Schwarz, M. Plenio, A. Retzker, L. P. McGuinness, and 
F. Jelezko, Submillihertz magnetic spectroscopy performed with
a nanoscale quantum sensor, Science \textbf{356}, 832 (2017).

\bibitem{boss2017quantum} J. M. Boss, K. S. Cujia, J. Zopes, and
C. L. Degen, Quantum sensing with arbitrary frequency resolution,
Science \textbf{356}, 837 (2017).

\bibitem{joas2017quantum}T. Joas, A. M. Waeber, G. Braunbeck, and
F. Reinhard, Quantum sensing of weak radio-frequency signals by pulsed
Mollow absorption spectroscopy, Nat. Commun. \textbf{8}, 964 (2017).

\bibitem{stark2017narrow}A. Stark, N. Aharon, T. Unden, D. Louzon,
A. Huck, A. Retzker, U. L. Andersen, and F. Jelezko, Narrow-bandwidth
sensing of high-frequency fields with continuous dynamical decoupling,
Nat. Commun. \textbf{8}, 1105 (2017).

\bibitem{chu2021precise} Y. Chu, P. Yang, M. Gong, M. Yu, B. Yu,
M. B. Plenio, A. Retzker, and J. Cai, Precise Spectroscopy of High-Frequency
Oscillating Fields with a Single-Qubit Sensor, Phys. Rev. Applied
\textbf{15}, 014031 (2021).

\bibitem{meinel2021heterodyne} J. Meinel, V. Vorobyov, B. Yavkin,
D. Dasari, H. Sumiya, S. Onoda, J. Isoya, and J. Wrachtrup, Heterodyne
sensing of microwaves with a quantum sensor, Nat. Commun. \textbf{12}, 1 (2021).

\bibitem{wang2022sensing} G. Wang, Y.-X. Liu, J. M. Schloss, S. T.
Alsid, D. A. Braje, and P. Cappellaro, Sensing of Arbitrary-Frequency
Fields Using a Quantum Mixer, Phys. Rev. X \textbf{12}, 021061 (2022).

\bibitem{jiang2023quantum} Z. Jiang, H. Cai, R. Cernansky, X. Liu,
and W. Gao, Quantum sensing of radio-frequency signal with NV centers
in SiC, Sci. Adv. \textbf{9}, eadg2080 (2023). 

%% ====Spin Control ==

\bibitem{taminiau2012detection} T. H. Taminiau, J. J. T. Wagenaar,
T. van der Sar, F. Jelezko, V. V. Dobrovitski, and R. Hanson, Detection
and Control of Individual Nuclear Spins using a Weakly Coupled Electron
Spin, Phys. Rev. Lett. \textbf{109}, 137602 (2012).


\bibitem{wang2017delayed} Z.-Y. Wang, J. Casanova, and M. B. Plenio,
Delayed entanglement echo for individual control of a large number
of nuclear spins, Nat. Commun. \textbf{8}, 14660 (2017).

\bibitem{haase2018soft} J. F. Haase, Z.-Y. Wang, J. Casanova, and
M. B. Plenio, Soft quantum control for highly selective interactions
among joint quantum systems, Phys. Rev. Lett. \textbf{121}, 050402 (2018). 

\bibitem{perlin2019noise} M. A. Perlin, Z.-Y. Wang, J. Casanova,
and M. B. Plenio, Noise-resilient architecture of a hybrid electron-nuclear
quantum register in diamond, Quantum Sci. Technol. \textbf{4}, 015007 (2019).

\bibitem{hegde2020efficient} S. S. Hegde, J. Zhang, and D. Suter,
Efficient quantum gates for individual nuclear spin qubits by indirect
control, Phys. Rev. Lett. \textbf{124}, 220501 (2020).

\bibitem{bartling2022entanglement} H. P. Bartling, M. H. Abobeih, B. Pingault, M.J. Degen, S. J. H. Loenen, C. E. Bradley, J. Randall, M. Markham, D. J. Twitchen, and T. H. Taminiau, Entanglement of Spin-Pair
Qubits with Intrinsic Dephasing Times Exceeding a Minute, Phys. Rev.
X \textbf{12}, 011048 (2022).

%%%

\bibitem{casanova2016noise} J. Casanova, Z.-Y. Wang, and M. B. Plenio,
Noise-Resilient Quantum Computing with a Nitrogen-Vacancy Center and
Nuclear Spins, Phys. Rev. Lett. \textbf{117}, 130502 (2016).

\bibitem{pezzagna2021quantum} S. Pezzagna and J. Meijer, Quantum
computer based on color centers in diamond, Appl. Phys. Rev. \textbf{8}, 011308
(2021).

%%%

\bibitem{cai2013a} J. Cai, A. Retzker, F. Jelezko, and M. B. Plenio,
A largescale quantum simulator on a diamond surface at room temperature,
Nat. Phys. \textbf{9}, 168 (2013).

%%  ===Network

\bibitem{childress2013diamond}L. Childress and R. Hanson, Diamond
NV centers for quantum computing and quantum networks, MRS Bull. \textbf{38},
134 (2013).

\bibitem{kalb2017entanglement}N. Kalb, A. A. Reiserer,
P. C. Humphreys, J. J. W. Bakermans,
S. J. Kamerling, N. H. Nickerson, S. C.
Benjamin, D. J. Twitchen, M. Markham, and R. Hanson,
Entanglement Distillation between Solid-State Quantum Network Nodes,
Science \textbf{356}, 928 (2017).

\bibitem{humphreys2018deterministic} P. C. Humphreys, N. Kalb, J.
P. J. Morits, R. N. Schouten, R. F. L. Vermeulen, D. J. Twitchen,
M. Markham, and R. Hanson, Deterministic delivery of remote entanglement
on a quantum network, Nature (London) \textbf{558}, 268 (2018).

\bibitem{pompili2021realization} M. Pompili, S. L. N.
Hermans, S. Baier, H. K. C. Beukers, P. C.
Humphreys, R. N. Schouten, R. F. L.
Vermeulen, M. J. Tiggelman, L. dos Santos Martins, B.
Dirkse, S. Wehner, and R. Hanson, Realization of a Multinode Quantum
Network of Remote Solid-State Qubits, Science \textbf{372}, 259 (2021).

%%%

\bibitem{carr1954effects} H. Y. Carr and E. M. Purcell, Effects of
diffusion on free precession in nuclear magnetic resonance experiments,
Phys. Rev. \textbf{94}, 630 (1954).

\bibitem{meiboom1958modified} S. Meiboom and D. Gill, Modified spin-echo
method for measuring nuclear relaxation times, Rev. Sci. Instrum.
\textbf{29}, 688 (1958).

\bibitem{gullion1990new} T. Gullion, D. B. Baker, and M. S. Conradi,
New, compensated Carr-Purcell sequences, J. Magn. Reson. \textbf{89}, 479 (1990).

%%% == Spurious ===


\bibitem{loretz2015spurious} M. Loretz, J. M. Boss, T. Rosskopf,
H. J. Mamin, D. Rugar, and C. L. Degen, Spurious Harmonic Response
of Multipulse Quantum Sensing Sequences, Phys. Rev. X \textbf{5}, 021009 (2015).

\bibitem{frey2017application} V. M. Frey, S. Mavadia, L. M. Norris,
W. de Ferranti, D. Lucarelli, L. Viola, and M. J. Biercuk, Application
of optimal band-limited control protocols to quantum noise sensing,
Nat. Commun. \textbf{8}, 2189 (2017).

\bibitem{zhao2014dynamical} N. Zhao, J. Wrachtrup, and R.-B. Liu,
Dynamical decoupling design for identifying weakly coupled nuclear
spins in a bath, Phys. Rev. A \textbf{90}, 032319 (2014).

\bibitem{casanova2015robust} J. Casanova, Z.-Y. Wang, J. F. Haase,
and M. B. Plenio, Robust dynamical decoupling sequences for individual
nuclear-spin addressing, Phys. Rev. A \textbf{92}, 042304 (2015).  

%\bibitem{key-95} J. F. Haase, Z.-Y. Wang, J. Casanova, and M. B.
%Plenio, Pulse-phase control for spectral disambiguation in quantum
%sensing protocols, Phys. Rev. A 94, 032322 (2016).
%
%\bibitem{key-91} Z. Shu, Z. Zhang, Q. Cao, P. Yang, M. B. Plenio, C. Müller, 
%J. Lang, N. Tomek, B. Naydenov, L. P. McGuinness, F. Jelezko, and 
%J. Cai, Unambiguous nuclear spin detection using an engineered
%quantum sensing sequence, Phys. Rev. A 96, 051402(R) (2017).

\bibitem{wang2019randomization} Z.-Y. Wang, J. E. Lang, S. Schmitt,
J. Lang, J. Casanova, L. McGuinness, T. S. Monteiro, F. Jelezko, and
M. B. Plenio, Randomization of Pulse Phases for Unambiguous and Robust
Quantum Sensing, Phys. Rev. Lett. \textbf{122}, 200403 (2019).

\bibitem{wang2020enhancing} Z.-Y. Wang, J. Casanova, and M. B. Plenio,
Enhancing the Robustness of Dynamical Decoupling Sequences with Correlated
Random Phases, Symmetry \textbf{12}, 730 (2020).


%%% Adiabatic Control

\bibitem{wang2016necessary} Z.-Y. Wang and M. B. Plenio, Necessary
and sufficient condition for quantum adiabatic evolution by unitary
control fields. Phys. Rev. A \textbf{93}, 052107 (2016).

\bibitem{xu2019breaking} K. Xu, T. Xie, F. Shi, Z.-Y. Wang, X. Xu,
P. Wang, Y. Wang, M. B. Plenio, and J. Du, Breaking the quantum adiabatic
speed limit by jumping along geodesics. Sci. Adv. \textbf{5}, eaax3800 (2019).

\bibitem{liu2022shortcuts} Y. Liu and Z.-Y. Wang, Shortcuts to adiabaticity
with inherent robustness and without auxiliary control, arXiv:2211.02543.

\bibitem{gong2023accelerated} M. Gong, M. Yu, R. Betzholz, Y. Chu, P. Yang, Z. Wang, and J. Cai, Accelerated quantum control in a three-level system by jumping along the geodesics, Phys. Rev. A \textbf{107}, L040602 (2023).

\bibitem{berry2009transitionless} M. V. Berry, Transitionless quantum
driving, J. Phys. A: Math. Theor. \textbf{42}, 365303 (2009).

\bibitem{odelin2019shortcuts} D. Géury-Odelin, A. Ruschhaupt, A. Kiely, E. Torrontegui, 
S. Marínez-Garaot, and J. G. Muga, Shortcuts
to adiabaticity: Concepts, methods, and applications,
Rev. Mod. Phys. \textbf{91}, 045001 (2019).


\bibitem{chruscinski2004geometric} D. Chruscinski, A. Jamiolkowski,
Geometric Phases in Classical and Quantum Mechanics (Birkhäuser, 2004).

\bibitem{see_SM} See Supplemental Material at {[}URL{]} for additional
information. 

%%%

\bibitem{cywinski2008how} \L . Cywi\'{n}ski, R. M. Lutchyn, C. P.
Nave, and S. Das Sarma, How to enhance dephasing time in superconducting
qubits, Phys. Rev. B \textbf{77}, 174509 (2008).

\bibitem{ryan2010robust} C. A. Ryan, J. S. Hodges, and D. G. Cory,
Robust Decoupling Techniques to Extend Quantum Coherence in Diamond,
Phys. Rev. Lett. \textbf{105}, 200402 (2010).

\bibitem{genov2017arbitrarily} G. T. Genov, D. Schraft, N. V. Vitanov,
and T. Halfmann, Arbitrarily Accurate Pulse Sequences for Robust Dynamical
Decoupling, Phys. Rev. Lett. \textbf{118}, 133202 (2017).


\bibitem{alvarez2008measuring} G. A. Alvarez and D. Suter, Measuring
the spectrum of colored noise by dynamical decoupling, Phys. Rev.
Lett. \textbf{107}, 230501 (2011).

\bibitem{tycko1984fixed} R. Tycko and A. Pines, Iterative schemes for
 broad-band and narrow-band population inversion in NMR, 
Chem. Phys. Lett. \textbf{111}, 462 (1984). 

%%%

%\bibitem{gardiner2004handbook} C. Gardiner, \textit{Handbook of Stochastic
%Methods for Physics, Chemistry, and the Natural Sciences} (Springer-Verlag,
%Berlin, 2004), Chap. 3.





\end{thebibliography}

\begin{thebibliography}{99}
\bibitem{sm:cai2012robust} J.-M. Cai, B. Naydenov, R. Pfeiffer, L.  P.
McGuinness, K.  D. Jahnke, F. Jelezklo, M.  B.
Plenio, and A. Retzker, Robust dynamical decoupling with concatenated
continuous driving, New J. Phys. 14, 113023 (2012).

\bibitem{sm:chu2021precise} Y. Chu, P. Yang, M. Gong, M. Yu, B. Yu,
M. B. Plenio, A. Retzker, and J. Cai, Precise Spectroscopy of High-Frequency
Oscillating Fields with a Single-Qubit Sensor, Phys. Rev. Applied
\textbf{15}, 014031 (2021).

\bibitem{sm:boss2017quantum} J. M. Boss, K. Cujia, J. Zopes, and
C. L. Degen, Quantum sensing with arbitrary frequency resolution,
Science 356, 83 (2017).

\bibitem{sm:schmitt2017submillihertz} S. Schmitt, T. Gefen, F. M. Stürner, T. Unden, G. Wolff, C. 
Müller, J. Scheuer, B. Naydenov, M. Markham, S. Pezzagna, J. 
Meijer, I. Schwarz, M. Plenio, A. Retzker, L. P. McGuinness, and 
F. Jelezko, Submillihertz magnetic spectroscopy performed with
a nanoscale quantum sensor, Science \textbf{356}, 832 (2017).

\bibitem{sm:gardiner2004handbook} C. Gardiner, Handbook of Stochastic
Methods for Physics, Chemistry, and the Natural Sciences (Springer-Verlag,
Berlin, 2004), Chap. 3

\bibitem{sm:wang2016necessary} Z.-Y. Wang and M. B. Plenio, Necessary
and sufficient condition for quantum adiabatic evolution by unitary
control fields. Phys. Rev. A 93, 052107 (2016).

\bibitem{sm:xu2019breaking} K. Xu, T. Xie, F. Shi, Z.-Y. Wang, X.
Xu, P. Wang, Y. Wang, M. B. Plenio, and J. Du, Breaking the quantum
adiabatic speed limit by jumping along geodesics. Sci. Adv. 5, eaax3800
(2019).

\bibitem{sm:liu2022shortcuts} Y. Liu and Z.-Y. Wang, Shortcuts to
adiabaticity with inherent robustness and without auxiliary control,
arXiv:2211.02543.

\bibitem{sm:wang2008an}{]} X. Wang, C.-S. Yu, and X. X. Yi, An alternative
quantum fidelity for mixed states of qudits, Phys. Lett. A 373, 58
(2008).
\end{thebibliography}
\end{document}